\documentclass[amsmath,amssymb,showpacs,twocolumn,superscriptaddress,pr]{revtex4-1}
\usepackage{graphicx,bm,color,subfigure}
\usepackage[T1]{fontenc}
\usepackage{mathrsfs}
\usepackage{bm}
\usepackage{amsmath}
\usepackage{amssymb}
\usepackage{graphicx}
\usepackage{esint}
\usepackage{multirow}
\usepackage{float}
\usepackage{array}
\usepackage{makecell}
\usepackage{harpoon}
\usepackage{booktabs}
\usepackage{gensymb}
\usepackage{simplewick}
\usepackage{subfigure}

\usepackage[english]{babel}

\begin{document}
\title{$g+ig$ topological superconductivity in the 30\degree-twisted bilayer graphene}
\author{Yu-Bo Liu}
\thanks{These two authors contributed equally to this work.}
\affiliation{School of Physics, Beijing Institute of Technology, Beijing 100081, China}
\author{Yongyou Zhang}
\thanks{These two authors contributed equally to this work.}
\affiliation{School of Physics, Beijing Institute of Technology, Beijing 100081, China}
\author{Wei-Qiang Chen}
\affiliation{Shenzhen Key Laboratory of Advanced Quantum Functional Materials and Devices, Southern University of Science and Technology, Shenzhen 518055, China}
\affiliation{Institute for Quantum Science and Engineering and Department of Physics, Southern University of Science and Technology, Shenzhen 518055, China}
\author{Fan Yang}
\email{yangfan_blg@bit.edu.cn}
\affiliation{School of Physics, Beijing Institute of Technology, Beijing 100081, China}
\date{\today}
\begin{abstract}
Based on our revised perturbational-band theory, we study possible pairing states driven by interaction in the electron-doped quasicrystal 30\degree-twisted bilayer graphene. Our mean-field study on the related t-J model predicts that, the beneath-van-Hove and beyond-van-Hove low doping regimes are covered by the chiral $d+id$ and $g+ig$ topological superconductivities (TSCs) respectively. The $g+ig$-TSC possesses a pairing angular momentum 4, and hence following each effective $C_{12}$- rotation by $\Delta\phi=n\pi/6$, the pairing phase changes $4\Delta\phi$. This intriguing TSC is novel, as it belongs to a special 2D $E_4$- irreducible representation of the effective $D_{12}$ point group unique to this quasicystal and absent on periodic lattices. The Ginzburg-Landau theory suggested that the $g+ig$- TSC originates from the Josephson coupling between the $d+id$ pairings on the two mono-layers.
\end{abstract}
\pacs{......}

\maketitle
%\tableofcontents
{\bf Introduction:}
The twisted multi-layer graphene family since synthesis have driven the revelation of such intriguing quantum phases \cite{caoyuan20181,caoyuan20182,Yankowitz2019,Dean2019,Yazdani2019,Efetov2019,David2019,Serlin2019,P_Kim2020,Zeldov2020,P_Kim2021,Caoyuan2021} as the superconductivity (SC), the correlated insulator, the nematic phase, the quantum anomalous Hall, etc, which have aroused great research interests \cite{Xu2018,Po2018,Yuan2018,YangFan2018,WuFeng20181,Kang2018,Isobe2018,Koshino2018,Fernandes2018,Kang2019,Po2019,Dai2019,Honerkamp20191,Gonzalez2019,Song2019,Ahn2019,Angeli2019,Lian2019,Jiangpeng2019,Vishwanath2019prl,Linyuping2019,MingXie2020,Abouelkomsan2020,Bultinck2020prx,Senthil2020,Liao2021,Chichinadze2020,Mohammad2020,Wuxianxin2020}. While most of these studies are focused on the low twist-angle regimes, here we consider the recently synthesized \cite{Ahn2018,Yao2018,Pezzini2020,Yan2019,Deng2020} 30\degree-twisted bilayer graphene (TBG). This quasi-crystal TBG (QC-TBG) system hosts no real-space period and is characterized by an exact dodecagonal rotation symmetry verified by various experiments\cite{Ahn2018,Yao2018,Pezzini2020,Yan2019,Deng2020}. While the single-particle properties of this material have been widely studied\cite{Ahn2018,Yao2018,Moon2013,Koshino2015,Moon2019,Park2019,Crosse2021,Yuanshengjun2019,Yuanshengjun20201,Yuanshengjun20202,Aragon2019}, physical properties driven by electron-electron (e-e) interaction have not been investigated. Here we focus on possible exotic pairing states in the doped QC-TBG.

It has been long to search SC in the graphene family. Particularly, various groups have proposed \cite{Doniach2007, Gonzalez2008, Honerkamp2008, Pathak2010, McChesney2010, Nandkishore2012, Wang2012, Kiesel2012,Honerkamp2014} that the chiral $d_{x^2-y^2}\pm id_{xy}$ (abbreviated as $d+id$ below) topological SC (TSC) can be driven in the mono-layer graphene by e-e interaction near the van Hove (VH) doping. Recently, this material has been successfully doped to the beyond-VH regime \cite{exp_VHS}, which puts on the agenda the search of the exotic $d+id$ chiral TSC. It's timely now to investigate what exotic pairing states can be induced by the Josephson coupling between the $d+id$ pairing order parameters in the two mono-layers in the QC-TBG. While such weak inter-layer Josephson coupling is not expected to significantly change the $T_c$, certain choice of the relative phase difference between the two pairing order parameters can lead to new irreducible representation (IRRP) of the enlarged point group in QC-TBG, and hence novel TSC. Generally, TSCs on periodic lattices with at most 6- folded rotation axes can be generated from 2D $E_1$ ($p+ip$) or $E_2$ ($d+id$) IRRPs of the point groups. However, the dodecagonal rotation symmetry in the QC-TBG allows for more 2D IRRPs and hence more novel TSCs with higher pairing angular momenta $L\ge3$.

On the other front, the SC on QCs has recently been a hot topic \cite{DeGottardi2013,YuPeng2013,Loring2016,Sakai2017,Andrade2019,Sakai2019,Varjas2019,Nagai2020,Caoye2020,Sakai2020,YYZhang2020,Hauck2021,Zhoubin2021}, particularly after the synthesis of the QC Al-Zn-Mg superconductor \cite{exp}. Presently, most of these studies are focused on the intrinsic QC such as the Penrose lattice. Instead, the QC-TBG belongs to extrinsic QC \cite{Moon2019} which hosts two perfect crystal layers with independent periods, wherein the quasi-periodic nature appears only in the perturbational coupling between the two layers.  This structural character enables us to treat the system in the $\mathbf{k}$- space via the perturbational band theory, which lends us more convenience and insight to understand the SC on the extrinsic QC. Besides, it's interesting to clarify the difference between the pairing natures on intrinsic and extrinsic QCs.

In this paper, we study the pairing states in the doped QC-TBG, adopting the t-J model on this QC. We revise the perturbational band theory to suit it to the study on the finite lattice involving e-e interaction. The mean-field (MF) pairing phase diagram in the low electron-doped regime shows that the beneath-VH and beyond-VH doping regimes are covered by the $d+id$ and $g+ig$ TSCs, respectively. The $g+ig$ TSC belongs to the 2D $E_4$- IRRP of the effective $D_{12}$ point group unique to this QC, which possesses a high $L=4$ and its gap phase changes $2\pi/3$ for each 30$\degree$- rotation. Such an intriguing TSC originates from the Josephson coupling between the $d+id$ pairings on the two monolayers, as suggested by the Ginzburg-Landau theory. Our results not only open the door to study e-e correlation driven phases on extrinsic QCs, but also reveal a novel TSC rare on periodic lattices.

{\bf Model and Approach:} We start from the following tight-binding (TB) model for the QC-TBG\cite{Moon2019},
\begin{equation}\label{H_TB}
H_{\text{TB}}=-\sum_{\mathbf{ij}\sigma}t_{\mathbf{ij}}c^{\dagger}_{\mathbf{i}\sigma}c_{\mathbf{j}\sigma},
\end{equation}
with $t_{\mathbf{ij}}$ provided in the SM\cite{SM} adopted from Ref. \cite{Moon2019}. Eq. (\ref{H_TB}) can be decomposed into the zeroth-order intra-layer hopping term $H_0$ and the perturbational inter-layer tunneling term $H'$ as
\begin{eqnarray}\label{H_perturb}
H_{0}&=&\sum_{\mathbf{k}\mu\alpha\sigma}c^{\dagger}_{\mathbf{k}\mu\alpha\sigma}c_{\mathbf{k}\mu\alpha\sigma}\varepsilon_{\mathbf{k}}^{\mu\alpha},\nonumber\\
H'&=&\sum_{\mathbf{kq}\alpha\beta\sigma}c^{\dagger}_{\mathbf{k}\text{t}\alpha\sigma}c_{\mathbf{q}\text{b}\beta\sigma}T^{\alpha\beta}_{\mathbf{kq}}+h.c.
\end{eqnarray}
Here $\mathbf{k}/\mathbf{q}$, $\mu(=\text{t(top)},\text{b(bottom)})$, $\alpha/\beta$ and $\sigma$ label the momentum, layer, band and spin respectively, $\varepsilon_{\mathbf{k}}^{\mu\alpha}$ is the monolayer dispersion and $T^{\alpha\beta}_{\mathbf{kq}}$ is given by\cite{MacDonald2011,Castro2007,Castro2012,Bistritzer2010,Moon2013,Koshino2015,Moon2019}
\begin{equation}\label{tunneling_coefficient}
T^{\alpha\beta}_{\mathbf{kq}}=\left\langle \mathbf{k}\alpha^{(\text{t})}\left|H_{\text{TB}}\right|\mathbf{q}\beta^{(\text{b})}\right\rangle.
\end{equation}
In thermal dynamic limit, the nonzero $T^{\alpha\beta}_{\mathbf{kq}}$ requires $\mathbf{k}+\mathbf{G}^{(\text{t})}=\mathbf{q}+\mathbf{G}^{(\text{b})}$\cite{Ahn2018,Yao2018,Moon2013,Koshino2015,Moon2019}, where $\mathbf{G}^{(\text{t/b})}$ represent the reciprocal lattice vectors for the t/b layers. Under this condition, we have
$T^{\alpha\beta}_{\mathbf{kq}}\propto t\left(\mathbf{k}+\mathbf{G}^{(\text{t})}\right)$, which decays promptly with $|\mathbf{k+G}^{(\text{t})}|$. Therefore each zeroth-order top-layer eigenstate $\left|\mathbf{k}\alpha^{(\text{t})}\right\rangle$ can only couple with a few isolated bottom-layer eigenstates $\left|\mathbf{q}\beta^{(\text{b})}\right\rangle$, and vice versa, which justifies the perturbational treatment.

Note that on our finite lattice with discrete momentum points, for each typical $\mathbf{k}$ on the top layer, no $\mathbf{q}$ on the bottom layer can make $\mathbf{k}+\mathbf{G}^{(\text{t})}=\mathbf{q}+\mathbf{G}^{(\text{b})}$ exactly satisfied. Therefore for each top-layer state $|\mathbf{k}\alpha^{(\text{t})}\rangle$, we abandon this relation and directly use Eq. (\ref{tunneling_coefficient}) to find the bottom-layer states $|\mathbf{q}_i\beta_i^{(\text{b})}\rangle$ which obviously couple with it. Then, for these $|\mathbf{q}_i\beta_i^{(\text{b})}\rangle$ states, we find again all the $|\mathbf{k}'_j\alpha_j^{(\text{t})}\rangle$ states which obviously couple with them. Gathering all these states related to $|\mathbf{k}\alpha^{(\text{t})}\rangle$ as bases to form a close sub-space, we can write down and diagonalize the Hamiltonian matrix in this sub-space to obtain all the eigen states. Among these states, the one having the largest overlap with $|\mathbf{k}\alpha^{(\text{t})}\rangle$ is marked as its perturbation-corrected state $|\widetilde{\mathbf{k}\alpha^{(\text{t})}}\rangle$, whose energy is marked as $\tilde{\varepsilon}^{\text{t}\alpha}_{\mathbf{k}}$. The procedure to get $|\widetilde{\mathbf{q}\beta^{(\text{b})}}\rangle$ and $\tilde{\varepsilon}^{\text{b}\beta}_{\mathbf{q}}$ is similar. For more details, see the Supplementary Material (SM)\cite{SM}. The approach developed here is a finite-lattice version of the second-order perturbational theory in Ref. \cite{Yao2018,Moon2013,Koshino2015}.
\begin{figure}[htbp]
\centering
\includegraphics[width=0.5\textwidth]{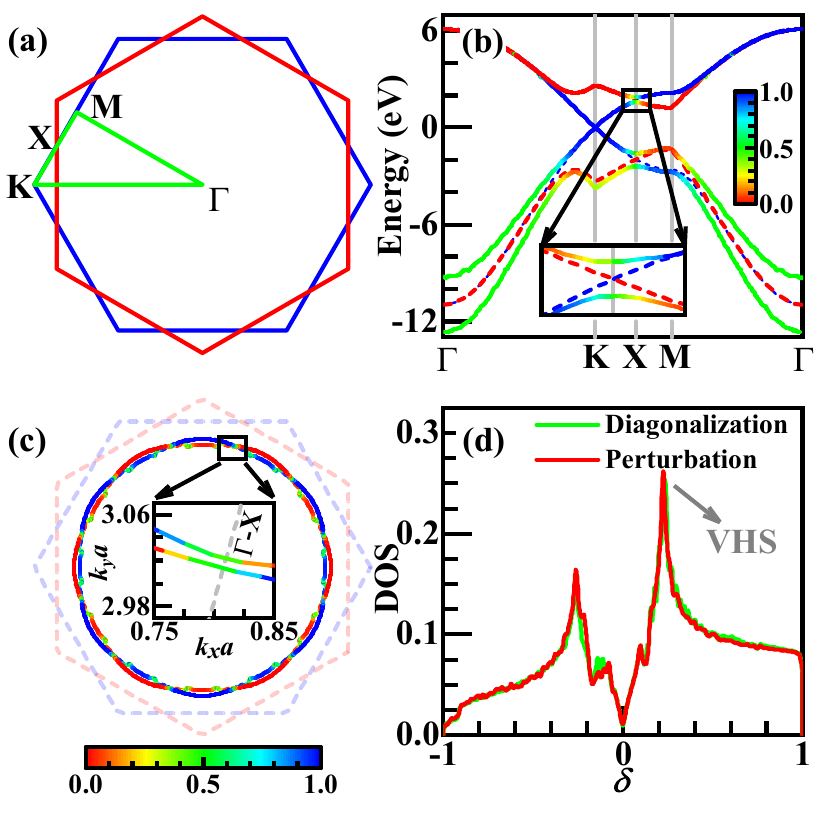}
\caption{(Color on line) (a) High-symmetry points marked in the Brillouin zone. (b) Band structure plotted along the high-symmetry lines: solid (dashed) lines for the QC-TBG (two uncoupled graphene monolayers). (c) FSs for the doping level $\delta=0.32$. (d) DOS from the perturbational band theory, in comparison with that from real-space diagonalization on a finite lattice with $9\times10^4$ sites. The color in (b, c) represents layer component. Insets in (b) band structure near the $X$ point and (c) FSs crossing the $\Gamma$-$X$ line (grey dotted line). }\label{band_structure}
\end{figure}

Along the lines connecting the high-symmetry points in the Brillouin zone marked in Fig. \ref{band_structure} (a), we plot our obtained band structure (solid lines) in Fig. \ref{band_structure} (b), in comparison with the uncoupled band structures (dashed lines) from the two layers. Remarkably, while there is obvious band splitting caused by inter-layer hybridization on the hole-doped side, the band structure on the electron-doped side is overall not far from simply overlaying the two uncoupled monolayer band structures, reflecting the much weaker inter-layer hybridization there\cite{Koshino2015,Moon2019}, proved in the SM\cite{SM}. Below, we shall focus on the electron-doped side as the weaker inter-layer coupling there validates our perturbational approach.

The main effect of the inter-layer coupling on the electron-doped side lies in that the top-layer band branches and the bottom-layer ones cross and split at the $X$ points, and simultaneously their layer components are exchanged (see the inset of Fig. \ref{band_structure} (b)). Actually, such band crossing and splitting takes place on all the $\Gamma$-$X$ lines: For each $\mathbf{k}$ on these lines, by symmetry, the states $|\mathbf{k}\alpha^{(\text{t})}\rangle$ and $|\mathbf{k}\alpha^{(\text{b})}\rangle$ possess degenerate zeroth-order energy. They are further coupled via $\mathbf{k}+\mathbf{G}^{(\text{t})}=\mathbf{k}+\mathbf{G}^{(\text{b})}$ by setting $\mathbf{G}^{(\text{t})}=\mathbf{G}^{(\text{b})}=0$, leading to their hybridization and the resulting band splitting. Consequently, the FSs contributed from the two layers also cross and split upon crossing these lines, and simultaneously the layer components are exchanged, see Fig. \ref{band_structure} (c). The emergent FSs with dodecagonal symmetry can be viewed as the inter-layer bonding- and anti-bonding- FSs.

The density of state (DOS) obtained by our approach is well consistent with that obtained by directly diagonalizing the real-space Hamiltonian, particularly in the electron-doped side, as shown in Fig. \ref{band_structure}(d). We have further checked that the obtained set of states $\{|\widetilde{\mathbf{k}\alpha^{(\mu)}}\rangle\}$ are almost mutually orthogonal. These properties well qualify them as a good set of single-particle bases for succeeding studies involving e-e interaction.

Now we come to the e-e interaction. As the local Hubbard $U$ for the graphene family can vary from 10 eV to 17 eV\cite{graphene_RMP}, comparable with the total band width $W_b\approx 18$ eV, the weak-coupling approach might not be suitable here. For simplicity, we take the strong-coupling limit, wherein the strong on-site repulsion induces the following superexchange interaction
\begin{equation}\label{superexchange}
H_J=\sum_{(\mathbf{i,j})}J_{\mathbf{ij}}\mathbf{S}_{\mathbf{i}}\cdot\mathbf{S}_{\mathbf{j}},
\end{equation}
with $J_{\mathbf{ij}}=4t_{\mathbf{ij}}^2/U$ and $U=10$ eV. Here the sum $\sum_{(\mathbf{i,j})}$ goes through the $(\mathbf{i,j})$ pairs on the bilayer. The total Hamiltonian is then given by the following t-J model,
\begin{equation}\label{total_Hamiltonian}
H=H_{\text{TB}}+H_J.
\end{equation}

Eq. (\ref{total_Hamiltonian}) is solved in the SM\cite{SM} by the MF or the Gutzwiller MF approaches, which yield the same pairing symmetry. Focusing on the intra-band pairing between opposite momenta, we obtain the pairing potential $V^{\mu\alpha}_{\nu\beta}(\mathbf{k},\mathbf{q})$ and the linearized gap equation near $T_c$,
\begin{eqnarray}\label{gap_equa}
V^{\mu\nu}_{\alpha\beta}&&(\mathbf{k},\mathbf{q})=\frac{-3 }{2N}\sum_{(\mathbf{i,j})}J_{\mathbf{ij}}\text{Re}\left(\tilde{\xi}_{\mathbf{i},\mathbf{k}\mu\alpha}\tilde{\xi}^*_{\mathbf{j},\mathbf{k}\mu\alpha}\right)
\text{Re}\left(\tilde{\xi}_{\mathbf{i},\mathbf{q}\nu\beta}\tilde{\xi}^*_{\mathbf{j},\mathbf{q}\nu\beta}\right),\nonumber\\
&&-\frac{1}{(2\pi)^2}\sum_{\nu\beta}\oint \frac{V^{\mu\nu}_{\alpha\beta}(\mathbf{k},\mathbf{q})}{v_F^{\nu\beta}(\mathbf{q})}\Delta_{\nu\beta}(\mathbf{q})dq_{\parallel}=\lambda\Delta_{\mu\alpha}(\mathbf{k}).
\end{eqnarray}
Here $\tilde{\xi}_{\mathbf{i},\mathbf{k}\mu\alpha}/\sqrt{N}$ denotes the wave function of $|\widetilde{\mathbf{k}\alpha^{(\mu)}}\rangle$, $v_F^{\nu\beta}(\mathbf{q})$ is the Fermi velocity and $q_{\parallel}$ denotes the component along the tangent of the FS. The pairing symmetry is determined by the gap form factor $\Delta_{\mu\alpha}(\mathbf{k})$ corresponding to the largest pairing eigenvalue $\lambda$ solved for Eq. (\ref{gap_equa}).

\begin{figure}[htbp]
\centering
\includegraphics[width=0.5\textwidth]{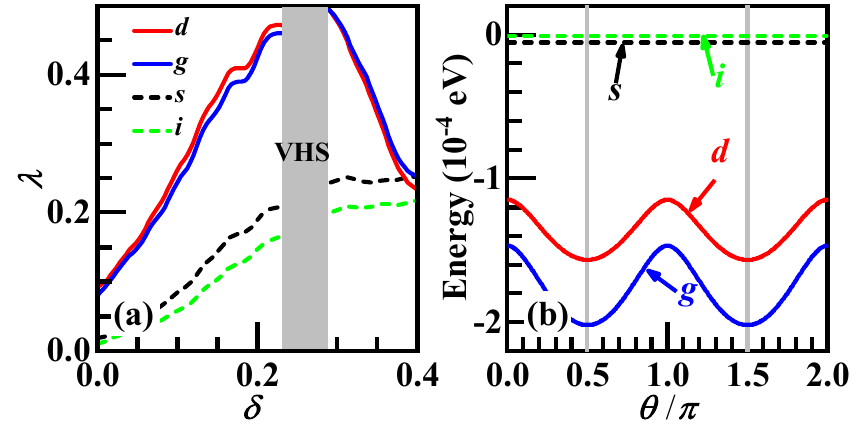}
\caption{(Color on line) (a)Doping $\delta$ dependences of the largest pairing eigenvalues $\lambda$ for the $s$-, $i$-, the degenerate $d$- and $g$- wave pairing symmetries for $\delta\in (0,0.4)$. The VH doping regime marked grey has been excluded. (b)Mixing-phase-angle $\theta$ dependences of the energies for the degenerate $d$- and $g$- wave pairings for $\delta=0.32$. The energies of the $s$- and $i$- wave pairings are also shown in comparison.}\label{phase_diagram}
\end{figure}

{\bf Phase diagram and Ginzburg-Landau theory:} The pairing symmetries can be classified according to the IRRPs of the point group $D_{6d}$, or effectively $D_{12}$ if we redefine the $C_{\frac{\pi}{6}}$ as the combination of it and the succeeding layer exchange. The $D_{12}$ possesses six singlet IRRPs and three triplet ones, see the SM\cite{SM}. The antiferromagnetic superexchange interaction here only allows for singlet pairings, including four 1D IRRPs: $s$, $i$, $i'$ and $i*i'$, and two 2D IRRPs: $d$ and $g$. The doping dependences of the largest $\lambda$ of these pairing symmetries within the experimentally accessible doping regime $\delta\in(0,0.4)$ are shown in Fig. \ref{phase_diagram}(a). Note that the regime very close to the VHS is excluded in this phase diagram, as the divergent DOS there might have led to other instabilities.

Figure \ref{phase_diagram}(a) shows that the beneath-VH and beyond-VH doping regimes are covered by the degenerate $(d_{x^2-y^2},d_{xy})$- and $(g_{x^4-6x^2y^2+y^4}, g_{x^3y-xy^3})$- doublets belonging to the 2D $E_2$- and $E_4$- IRRPs, respectively. The $s$-wave SC can emerge for $\delta>0.4$, and the other symmetries including the shown $i$-wave are always suppressed. The gap functions of these pairing states are provided in the SM\cite{SM}. The two components of the $d$- or $g$-wave pairings are mixed as $1:\alpha e^{i\theta}$, and consequently the ground-state energies shown in Fig. \ref{phase_diagram}(b) are minimized at $\alpha=1$ and $\theta=\pm\pi/2$\cite{SM}, leading to the $d+id$- or $g+ig$- SCs, which can be well understood by the G-L theory\cite{SM}.

Note that each possible ground-state gap function $\Delta(\mathbf{k})$ forms a 1D IRRP of the $C_{12}$ sub-group\cite{SM}: for each rotation by the angle $\Delta\phi=n\pi/6$, we have $\Delta(\mathbf{k})\to e^{-i\Delta\phi_{SC}}\Delta(\mathbf{k})= e^{-iL\Delta\phi}\Delta(\mathbf{k})$, with $L=0$ for the $s$ and $i*i'$, $L=2$ for the $d+id$, $L=4$ for the $g+ig$ and $L=6$ for the $i$ and $i'$. Fig. \ref{phase_diagram} suggests that the cases $L=2,4$ are realized in the QC-TBG. Interestingly, this result can be understood by the following G-L theory from the weak Josephson coupling between the previously obtained $d+id$ pairings in the two monolayers \cite{Doniach2007,Gonzalez2008,Honerkamp2008,Pathak2010,McChesney2010,Nandkishore2012,Wang2012,Kiesel2012,Honerkamp2014}.

Suppose the normalized gap form factors (either in the real or $\mathbf{k}$ space) on the t/b layers are $\Delta^{(\text{t/b})}_{d+id}$, we have\cite{footnote2}
\begin{equation}\label{relation_gap_updn}
\Delta^{(\text{b})}_{d+id}=\hat{P}_{\frac{\pi}{6}}\Delta^{(\text{t})}_{d+id},~~ \hat{P}_{\frac{\pi}{3}}\Delta^{(\mu)}_{d+id}=e^{-i\frac{2\pi}{3}}\Delta^{(\mu)}_{d+id}.
\end{equation}
Here $\hat{P}_{\phi}$ denotes the rotation by $\phi$. Setting the global ``complex amplitudes'' of the pairing gap functions on the t/b layers as $\psi_{\text{t/b}}$ \cite{footnote3}, the free energy function $F$ satisfying all the symmetries of QC-TBG reads (see the SM\cite{SM}),
\begin{eqnarray}\label{G_L_F}
F\left(\psi_{\text{t}},\psi_{\text{b}}\right)=&F&_0(\left|\psi_{\text{t}}\right|^2)+F_0(\left|\psi_{\text{b}}\right|^2)\nonumber\\&-&A\left(e^{i\theta}\psi_{\text{t}}\psi_{\text{b}}^*+c.c\right)+O\left(\psi^4\right).
\end{eqnarray}
Here the $F_0$- and the $A$- terms denote the contributions from each monolayer and their Josephson coupling respectively. The $A>0$ and $\theta\in(-\pi,\pi]$ are real numbers.
\begin{figure}[htbp]
\centering
\includegraphics[width=0.5\textwidth]{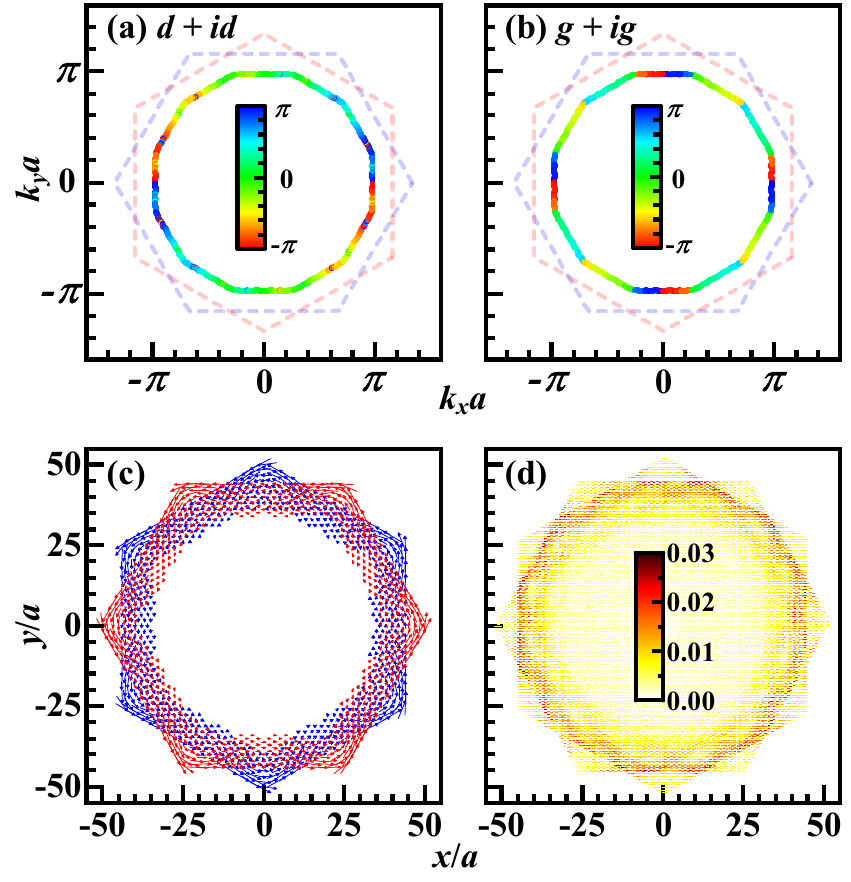}
\caption{(Color on line) The distributions of the gap phases for the $d+id$- (a) and the $g+ig$- (b) TSCs within a narrow energy shell near the FS. The real-space distributions of the spontaneous super current (c) and a typical Majorana zero mode (d). The doping level is $\delta=0.32$.}\label{topology}
\end{figure}

Now let's rotate the system by $\Delta\phi=\frac{\pi}{6}$, followed by a succeeding layer exchange, under which Eq. (\ref{relation_gap_updn}) dictates
\begin{equation}\label{ODP_change}
\psi_\text{b}\to\tilde{\psi_\text{b}}=\psi_\text{t}, \psi_\text{t}\to\tilde{\psi_\text{t}}=e^{-i\frac{2\pi}{3}}\psi_\text{b}.
\end{equation}
Here we use the convention with fixed $\Delta^{(\mu)}_{d+id}$. By symmetry, we have $F(\tilde{\psi_\text{t}},\tilde{\psi_\text{b}})=F(\psi_\text{t},\psi_\text{b})$, dictating
\begin{equation}
e^{i(\theta-2\pi/3)}=e^{-i\theta}\Rightarrow\theta=\frac{\pi}{3} \text{~or~} -\frac{2\pi}{3}
\end{equation}
Under $A>0$, as the two layers are identical, $F$ should be minimized at $\psi_\text{b}=e^{i\theta}\psi_\text{t}$. The solution $\theta=\pi/3$ or $-2\pi/3$ makes $(\tilde{\psi_\text{t}}, \tilde{\psi_\text{b}})=e^{-i\frac{\pi}{3}}(\psi_\text{t},\psi_\text{b})$ or $e^{i\frac{2\pi}{3}}(\psi_\text{t},\psi_\text{b})$, corresponding to $L=\frac{\pi}{3}/\Delta\phi=2$ or $\frac{-2\pi}{3}/\Delta\phi=-4$, respectively. Both solutions are supported by our microscopic calculations in different doping regimes.

More heuristically, our results can also be understood directly by the definition $L=\Delta\phi_{SC}/\Delta\phi$. As there is a $2n\pi$ uncertainty in $\Delta\phi_{SC}$, we should investigate the rotation by the minimum angle $\Delta\phi_m$. For each monolayer with $D_6$ symmetry, $\Delta\phi_m=\pi/3$. If we rotate the QC-TBG by $\pi/3$, then the $d+id$ pairing with $L=2$ within each mono-layer causes $\Delta \phi_{SC}=2\pi/3$ in that layer, as well as in the coupled system. This $\Delta \phi_{SC}$ can also be indistinguishably viewed as $-4\pi/3$. However, the emergent $D_{12}$ symmetry in the QC-TBG allows for the rotation by $\Delta\phi_m=\pi/6$, under which we have either $\Delta \phi_{SC}=\pi/3$ causing $L=2$, or $\Delta \phi_{SC}=-2\pi/3$ causing $L=-4$, which are distinguishable now. Our microscopic calculation results allow for both phases in different doping regimes.

{\bf Novel TSC:} The distributions of the gap amplitudes of the obtained $d+id$- and $g+ig$- SCs on the FSs are fully gapped\cite{SM}, and those of their gap phases are shown in Figs. \ref{topology} (a) and (b) on the inner pocket for $\delta=0.32$ (those on the outer pocket are shown in the SM\cite{SM}). Clearly for each run around the pocket, the gap phases of the $d+id$- and $g+ig$- TSCs changes $4\pi$ and $8\pi$, leading to the winding numbers $2$ and $4$, and hence the Chern- numbers $C=2$ and $4$ per pocket respectively\cite{SM}.

The $g+ig$- TSC obtained here is novel, as it belongs to the 2D $E_4$- IRRP absent for the typical $D_{2n}$ point group on periodic lattices. Generally, the 2D IRRPs of $D_{2n}$ include the $E_L$ with $L\le n-1$\cite{SM}. For periodic lattices, we have $n\le 3$ and hence $L\le 2$, corresponding to the $p+ip$ and $d+id$. Here in the QC-TBG, we have $n=6$ and hence $L\le 5$, including the $g+ig$. Although on periodic lattices with, say $D_6$ symmetry, the $g+ig$ can also emerge as a higher-harmonics basis function of the $E_2$- IRRP, it's generally mixed with the $d+id$ as both belong to the $E_2$. Occasionally, the mixing weight of the $d+id$- component vanishes and one gets the $g+ig$\cite{Chichinadze2020}, which is accidental and rare. Here, the $g+ig$ TSC is protected by symmetry not to mix with the $d+id$ one.

The difference between the properties of the TSCs on extrinsic and intrinsic QCs is clarified here. It's illustrated in Ref. \cite{Caoye2020} that on such intrinsic QC as the Penrose lattice, the lack of translational symmetry causes spontaneous bulk super current in chiral TSCs. However, as shown in Fig. \ref{topology}(c) for the chiral $g+ig$ TSC obtained with open boundary condition, the super current only distributes at the edge with chiral pattern. As proved in the SM\cite{SM}, the bulk current vanishes in the case of intra-band pairing between opposite momenta, caused by the nearly translational symmetry here. Similarly, the Majorana zero modes\cite{SM} are also localized at the edge, as shown in Fig. \ref{topology}(d). These properties make the TSCs on extrinsic QCs different from those on intrinsic QCs.

{\bf Conclusion and Discussions:} In conclusion, we have revised the perturbational band theory to suit it for the study of e-e interaction driven SCs on extrinsic QCs. This $\mathbf{k}$-space approach is more convenient than the real-space ones\cite{Loring2016,Sakai2017,Andrade2019,Sakai2019,Nagai2020,Caoye2020,Sakai2020,YYZhang2020} to study the properties of the pairing states, as we can intuitively examine the distribution of the pairing gap functions on the FS. Consequently, this work for the QC-TBG reveals a novel $g+ig$ TSC with $L=4$ rare on periodic lattices, whose physical origin can be well understood from the G-L theory.

This approach also applies to other extrinsic QCs. One example is the 45\degree-twisted bilayer cuprate film. While Ref. \cite{cuprates_QC} has studied the cases with commensurate twist angle $\theta$, the case of $\theta=45\degree$ can be studied by our approach. This approach can also be used to study other instabilities. One example is the spin density wave (SDW) in the VH-doped QC-TBG induced by the weak coupling between the chiral SDW orders\cite{Wang2012, Li2012} in the two mono-layers. Finally, since our approach is fully microscopic, we can use it to control electron phases through tuning such parameters as the inter-layer coupling or the displacement field, to achieve more exotic phases.

\section*{Acknowledgements}
We acknowledge stimulating discussions with Cheng-Cheng Liu, Ye Cao, Zheng-Cheng Gu and Yan-Xia Xing. This work is supported by the NSFC under the Grant Nos. 12074031, 12074037, 11674025,11861161001. W.-Q. Chen is supported by the Science, Technology and Innovation Commission of Shenzhen Municipality (No. ZDSYS20190902092905285), Guangdong Basic and Applied Basic Research Foundation under Grant No. 2020B1515120100 and Center for Computational Science and Engineering of Southern University of Science and Technology.

\newpage

\renewcommand{\theequation}{S\arabic{equation}}
\setcounter{equation}{0}
\renewcommand{\thefigure}{S\arabic{figure}}
\setcounter{figure}{0}
\renewcommand{\thetable}{S\arabic{table}}
\setcounter{table}{0}
\begin{widetext}
\section{\label{sec:level1}perturbational band theory }

This section provides some details of the perturbational-band-theory approach adopted in our study for the free-electron part of the quasi-crystal twisted bilayer graphene (QC-TBG). We will provide here detailed information of the band structure, the Fermi surface (FS) and density of state (DOS) thus obtained for the system.

We start from the following tight-binding (TB) model of the QC-TBG,
\begin{equation}\label{H_TB}
H_{\text{TB}}=-\sum_{\mathbf{ij}\sigma}t_{\mathbf{ij}}c^{\dagger}_{\mathbf{i}\sigma}c_{\mathbf{j}\sigma},
\end{equation}
where $\sigma$ labels spin and the hopping integral $t_{\mathbf{ij}}$ between site $\mathbf{i}$ and site $\mathbf{j}$ is given as \cite{1}
\begin{equation}
-t_{\mathbf{ij}}=t_{\mathbf{ij}\pi}\left [1-\left(\frac{\mathbf{R}_{\mathbf{ij}}\cdot\mathbf{e}_{\mathbf{z}}}{R}\right)^{2}\right]+
t_{\mathbf{ij}\sigma}\left(\frac{\mathbf{R}_{\mathbf{ij}}\cdot\mathbf{e}_{\mathbf{z}}}{R}\right)^{2},
\end{equation}
with
\begin{equation}
t_{\mathbf{ij}\pi}=t_{\pi}e^{-\left(R_{\mathbf{ij}}-a/\sqrt{3}\right)/r_{0}},\quad
t_{\mathbf{ij}\sigma}=t_{\sigma}e^{-(R_{\mathbf{ij}}-d)/r_{0}}.\nonumber
\end{equation}
Here, $R_\mathbf{ij}$ is the length of the vector $\mathbf{R}_\mathbf{ij}$, pointing from site $\mathbf{i}$ to site $\mathbf{j}$, and $\mathbf{e}_{\mathbf{z}}$ is the unit vector perpendicular to the graphene layer. The lattice constant and interlayer spacing are given by $a\approx0.246$ nm and $d\approx0.335$ nm, respectively. The parameters $t_{\pi}\approx-2.7$ eV, $t_\sigma\approx0.48$ eV and $r_{0}\approx0.0453$ nm are taken according to Ref.~\cite{1}.

The Hamiltonian \eqref{H_TB} could be decomposed into the zeroth-order intralayer hopping term $H_0$ and perturbational inter-layer tunneling term $H'$, namely,
\begin{eqnarray}\label{perturbation}
H_0=\sum_{\textbf{k}\mu\alpha\sigma}  c_{\textbf{k}\mu\alpha\sigma}^\dag c_{\textbf{k}\mu\alpha\sigma}\varepsilon_{\textbf{k}}^{\mu\alpha},\qquad
H'=\sum_{\textbf{kq}\alpha\beta\sigma}  c_{\textbf{k}{\rm t}\alpha\sigma}^\dag c_{\textbf{q}{\rm b}\beta\sigma}T_{\textbf{kq}}^{\alpha\beta}+h.c.,
\end{eqnarray}
where $\mu\ [=\rm \text{t (top)},\text{b (bottom)}]$ and $\alpha\ (=1,2)$ are the layer and band indices, respectively, $\textbf{k}$ and $\textbf{q}$ label the momentum, and $\varepsilon_{\textbf{k}}^{\mu\alpha}$ is the dispersion of single-layer graphene. The eigenstate of $H_0$ is denoted as $| \textbf{k}\alpha^{(\mu)}\rangle$ (here we omit the spin index for simplicity, since the following treatment has nothing to do with the spin degree of freedom), representing a monolayer state on the layer $\mu$. The interlayer tunneling matrix element $T_{\textbf{kq}}^{\alpha\beta}$ reads
\begin{eqnarray}    \label{interlayer_hopping}
T_{\textbf{kq}}^{\alpha\beta}=\langle \textbf{k}\alpha^{(\rm t)}|H_{\text{TB}} |\textbf{q}\beta^{(\rm b)}\rangle
= -\frac{1}{N}\sum_{\textbf{ij}}
\xi_{\textbf{i},\textbf{k}{\rm t}\alpha}^{*}
\xi_{\textbf{j},\textbf{q}{\rm b}\beta}
t_{\mathbf{ij}},
\end{eqnarray}
where $\xi/\sqrt{N}$ ($N$ is the number of unit cells on each monolayer, i.e. $\frac{1}{4}$ of the total site number) represents the real-space wave function for the monolayer state $| \textbf{k}\alpha^{(\mu)}\rangle$.

Our perturbational-band theory is a numerical version of the usual analytical second-order perturbational theory. Given a zeroth-order state $|\mathbf{k}\alpha^{(\mu)}\rangle$  with the zeroth-order energy $\varepsilon_{\textbf{k}}^{\mu\alpha}$, we provide in the following our procedure to obtain its perturbation-corrected state $|\widetilde{\mathbf{k}\alpha^{(\mu)}}\rangle$, and the perturbation-corrected energy $\tilde{\varepsilon}^{\mu\alpha}_{\mathbf{k}}$. For a state $| \textbf{k}\alpha^{(\rm t)}\rangle$ in the top layer, one can find some states $|\textbf{q}\beta^{(\rm b)}\rangle$ in the bottom layer which couple with it through Eq. (\ref{interlayer_hopping}). In thermal dynamic limit, the nonzero coupling matrix element $T^{\alpha\beta}_{\mathbf{kq}}$ requires~\cite{1,2,3}
\begin{equation}\label{couple_condition}
\mathbf{k}+\mathbf{G}^{(\text{t})}=\mathbf{q}+\mathbf{G}^{(\text{b})},
\end{equation}
where $\mathbf{G}^{(\text{t/b})}$ represent the reciprocal lattice vectors of the top/bottom layers. However, on our finite lattice with discrete momentum points, for each typical $\mathbf{k}$ on the top layer, no $\mathbf{q}$ on the bottom layer can make the relation (\ref{couple_condition}) exactly satisfied unless $\mathbf{k}=\mathbf{q}=\mathbf{G}^{(\text{t})}=\mathbf{G}^{(\text{b})}=0$, as the two sets of relatively $30\degree$ rotated lattices are mutually incommensurate with each other. Therefore for each $|\mathbf{k}\alpha^{(\text{t})}\rangle$ state on the top layer, we abandon this relation and directly use Eq. (\ref{interlayer_hopping}) to numerically find the $|\mathbf{q}_i\beta_i^{(\text{b})}\rangle$ states on the bottom layer which obviously couple with this state. Here we only keep those $|\mathbf{q}_i\beta_i^{(\text{b})}\rangle$ states when their tunneling strengths $T_{\textbf{kq}}^{\alpha\beta}$ with $|\mathbf{k}\alpha^{(\text{t})}\rangle$ are more than $0.2$ times of the maximum one in all $T_{\textbf{kq}}^{\alpha\beta}$. One can imagine that the momenta of these kept states only make the relation (\ref{couple_condition}) nearly satisfied. Then, for these $|\mathbf{q}_i\beta_i^{(\text{b})}\rangle$ states, we find again all the $|\mathbf{k}'_j\alpha_j^{(\text{t})}\rangle$ states on the top layer which obviously couple with them. Gathering all these states related to $|\mathbf{k}\alpha^{(\text{t})}\rangle$ as bases to form a close sub-space, we can write down and diagonalize the Hamiltonian matrix in this sub-space to obtain all the eigen states. Among these states, the one having the largest overlap with $|\mathbf{k}\alpha^{(\text{t})}\rangle$ is marked as its perturbation-corrected state $|\widetilde{\mathbf{k}\alpha^{(\text{t})}}\rangle$, whose energy is marked as $\tilde{\varepsilon}^{\text{t}\alpha}_{\mathbf{k}}$. The procedure to get $|\widetilde{\mathbf{q}\beta^{(\text{b})}}\rangle$ and $\tilde{\varepsilon}^{\text{b}\beta}_{\mathbf{q}}$ is similar.

\begin{figure}[htbp]
	\centering
	\includegraphics[width=0.95\textwidth]{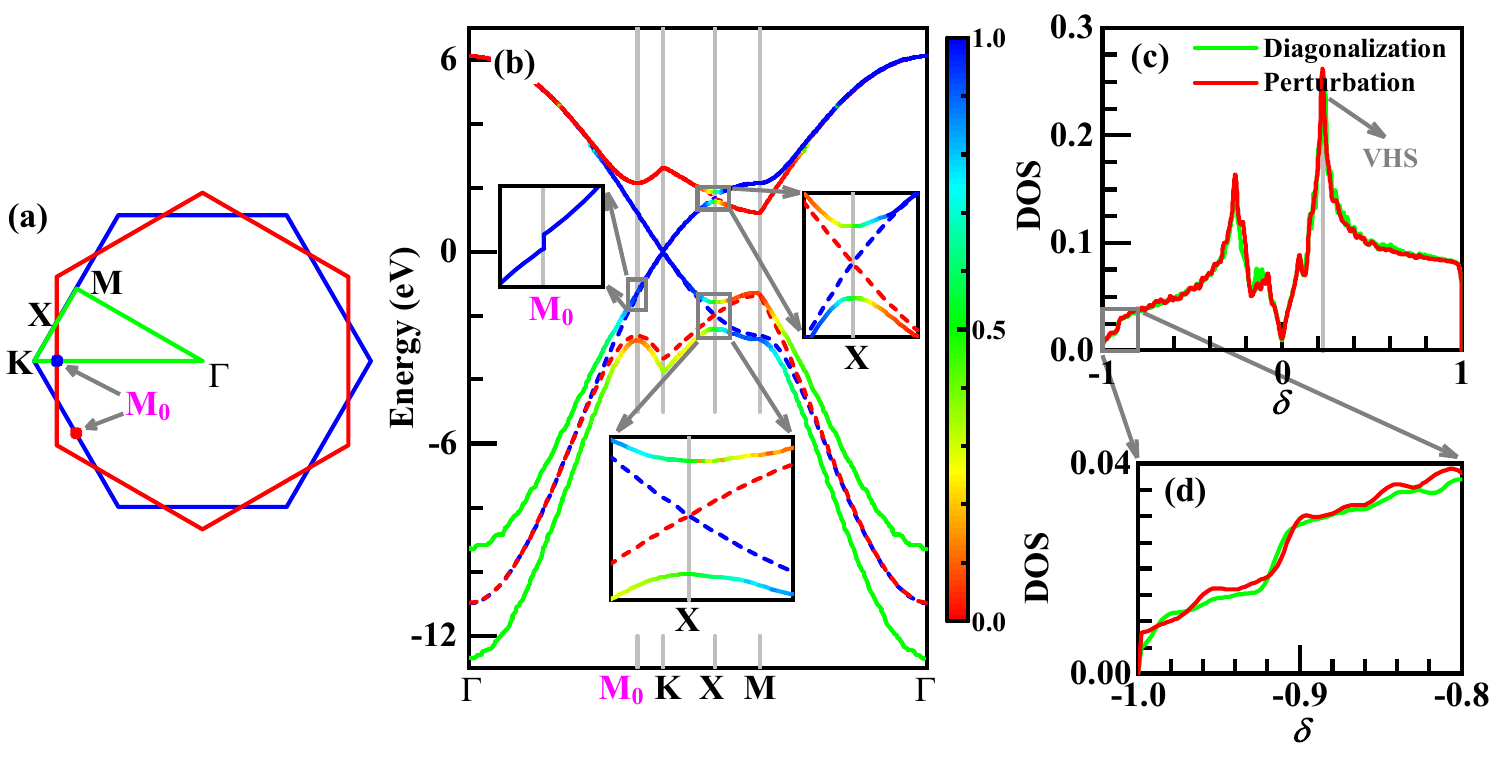}
	\caption{(a) High-symmetry points marked in the Brillouin zone. (b) Band structures of the QC-TBG with inter-layer tunneling (solid lines) and without inter-layer tunneling (dashed lines) along the high-symmetry lines in (a). (c) DOS calculated by the perturbational-band theory in comparison with that obtained by real-space diagonalization on a finite lattice with 90000 sites. (d) Enlarged view of the DOS around $\delta=-0.9$, presenting the twice step-like drops of the DOS.}
	\label{dos-gap}
\end{figure}
Along the lines connecting the high-symmetry points in the Brillouin zone marked in Fig.~\ref{dos-gap}(a), we plot our obtained band structure (solid lines) in Fig.~\ref{dos-gap}(b), in comparison with the uncoupled band structures (dashed lines) from the two layers. A remarkable feature of  Fig.~\ref{dos-gap}(b) is the obvious particle-hole (p-h) asymmetry: while the band structure on the electron-doped side is overall not far from simply overlaying the two sets of uncoupled monolayer band structures, there is strong interlayer hybridization and split within the band-bottom regime near the $\Gamma$-point on the hole-doped side. This split is reflected by the twice step-like drops in the density of states at the lower filling region (with doping level $\delta\approx-0.9$) in the hole-doped side, as shown in Figs.~\ref{dos-gap}(c) and \ref{dos-gap}(d), which is different from the monolayer graphene and is consistent with the result from  Ref.~\cite{3}. Such a p-h asymmetry is caused by the relatively weaker interlayer coupling on the electron-doped side, as was revealed in Refs.~\cite{1,3} and proved in the following.

Near the $\Gamma$ point, the band structure on each layer comprises two bands: one band labeled as ``$+$'' originates from the bonding between the A and B sublattices, while the other labeled as ``$-$'' originates from the anti-bonding between the two sublattices. The formula of the four zeroth-order states near the $\Gamma$ point for the two layers are as following,
\begin{eqnarray}\label{couplings}
|\textbf{k}+^{(\text{t})}\rangle&\approx&\frac{1}{\sqrt{2}}\left(|\textbf{k}A^{(\text{t})}\rangle+|\textbf{k}B^{(\text{t})}\rangle\right),\qquad
|\textbf{k}-^{(\text{t})}\rangle\approx\frac{1}{\sqrt{2}}\left(|\textbf{k}A^{(\text{t})}\rangle-|\textbf{k}B^{(\text{t})}\rangle\right),\nonumber\\
|\textbf{q}+^{(\text{b})}\rangle&\approx&\frac{1}{\sqrt{2}}\left(|\textbf{q}A^{(\text{b})}\rangle+|\textbf{q}B^{(\text{b})}\rangle\right),\qquad
|\textbf{q}-^{(\text{b})}\rangle\approx\frac{1}{\sqrt{2}}\left(|\textbf{q}A^{(\text{b})}\rangle-|\textbf{q}B^{(\text{b})}\rangle\right).
\end{eqnarray}
On each layer, the energy of the state labeled by ``$+$'' is lower than that of the state labeled by ``$-$''. Therefore, the former occupies the band bottom regime and the latter occupies the band top regime. Then we consider the interlayer couplings between each two zeroth-order states from different layers.  Before that, we first evaluate the coupling between the states $|\textbf{k}X^{{(\rm t)}}\rangle$ and $|\textbf{q}Y^{(\rm b)}\rangle$~\cite{1} [here $X$ and $ Y (=\rm A,\ B)$ are sublattice indices],
 \begin{eqnarray} \label{ss6}
 \langle \textbf{k}X^{{(\rm t)}}|H_{\text{TB}} |\textbf{q}Y^{(\rm b)}\rangle=-\sum_{\mathbf{G}^{(\rm t)}\mathbf{G}^{(\rm b)}}t(\mathbf{k}+\mathbf{G}^{(\rm t)})e^{i\mathbf{G}^{(\rm t)}\mathbf{\tau}_{X}-i\mathbf{G}^{(\rm b)}\mathbf{\tau}_{Y}}\delta_{\mathbf{k}+\mathbf{G}^{(\rm t)},\mathbf{q}+\mathbf{G}^{(\rm b)}}
 \end{eqnarray}
 with
 \begin{eqnarray}
 t(\mathbf{k}+\mathbf{G}^{(\rm t)})=\frac{1}{N}\sum_{\mathbf{i}\mathbf{j}} t_{\mathbf{i}\mathbf{j}} e^{-i(\mathbf{k}+\mathbf{G}^{(\rm t)})\mathbf{R}_{\mathbf{i}\mathbf{j}}} \label{tkG}
 \end{eqnarray}
where $\mathbf{\tau}$ is the sublattice position. Since $ t(\mathbf{k}+\mathbf{G}^{(\rm t)})$ decays exponentially for large $\mathbf{k}+\mathbf{G}^{(\rm t)}$, we only consider $\textbf{G}^{(\rm t)}=0$. Further more, the nonzero value of the $\delta_{\mathbf{k}+\mathbf{G}^{(\rm t)},\mathbf{q}+\mathbf{G}^{(\rm b)}}$ function requires $\textbf{G}^{(\rm b)}=0$ for $\mathbf{k}/\mathbf{q}\approx 0$ near the $\Gamma$ point. Then, equation \eqref{ss6}  can be written as
 \begin{eqnarray}\label{matrix_element_AB}
 \langle \textbf{k}X^{{(\rm t)}}|H_{\text{TB}} |\textbf{q}Y^{(\rm b)}\rangle\approx-t(\mathbf{k})\delta_{\mathbf{k},\mathbf{q}}.
 \end{eqnarray}
From Eq.~\eqref{couplings} and Eq.~\eqref{matrix_element_AB}, one gets
\begin{eqnarray}
&\langle\textbf{k}+^{{(\rm t)}}|H_{\text{TB}} |\textbf{q}+^{(\rm b)}\rangle \approx -2t(\mathbf{k})\delta_{\mathbf{k},\mathbf{q}}, \qquad
&\langle\textbf{k}-^{(\rm t)}|H_{\rm TB}|\textbf{q}-^{(\rm b)}\rangle \approx 0,\nonumber \\
&\langle\textbf{k}+^{(\rm t)}|H_{\rm TB}|\textbf{q}-^{(\rm b)}\rangle \approx 0, \qquad\qquad\qquad\
&\langle\textbf{k}-^{(\rm t)}|H_{\rm TB}|\textbf{q}+^{(\rm b)}\rangle \approx 0.
\end{eqnarray}
Therefore, one can see that the strong interlayer coupling only takes place in the band bottom regime of the band structure. This explains the p-h asymmetry character of the band structure.
Below, we shall focus on the electron-doped side as the weaker interlayer coupling there validates our perturbational approach.

The main effect of the interlayer coupling on the electron-doped side lies in that the top-layer band branches and the bottom-layer ones cross and split at the $X$-point, after which they exchange their layer components, as shown in the inset of Fig.~\ref{dos-gap}(b). Actually, such band crossing and splitting take place on the whole $\Gamma$-$X$ line: for each $\mathbf{k}$ on this line, by symmetry, the states $|\mathbf{k}\alpha^{(\text{t})}\rangle$ and $|\mathbf{k}\alpha^{(\text{b})}\rangle$ possess degenerate zeroth-order energy. They are further coupled via $\mathbf{k}+\mathbf{G}^{(\text{t})}=\mathbf{k}+\mathbf{G}^{(\text{b})}$ by setting $\mathbf{G}^{(\text{t})}=\mathbf{G}^{(\text{b})}=0$. The perturbational coupling between the two degenerate states $|\mathbf{k}\alpha^{(\text{t})}\rangle$ and $|\mathbf{k}\alpha^{(\text{b})}\rangle$ leads to their hybridization into bonding and anti-bonding states with an energy split between each other, causing the band splitting. Consequently, the FSs contributed from the two layers also cross and split upon crossing that line, after which they exchange their layer components, as shown in the insets of Fig.~\ref{fs} for different dopings. As a consequence of this interlayer coupling, the emergent bonding and anti-bonding FSs possess dodecagonal symmetry. Besides, our calculation reveals tiny gaps (about 0.1eV) at the points $\mathbf{M}_{0}^{(\rm t/b)}=\mathbf{G}^{(\rm b/t)}/2$ shown in the inset of Fig.~\ref{dos-gap}(b). These gaps are caused by the second-order perturbational coupling between the states $\left|\mathbf{M}_{0}^{(\mu)}\alpha^{(\mu)}\right\rangle$ and $\left|-\mathbf{M}_{0}^{(\mu)}\alpha^{(\mu)}\right\rangle$, consistent with Ref.~\cite{2}.

\begin{figure}[htbp]
	\centering
	\includegraphics[width=0.95\textwidth]{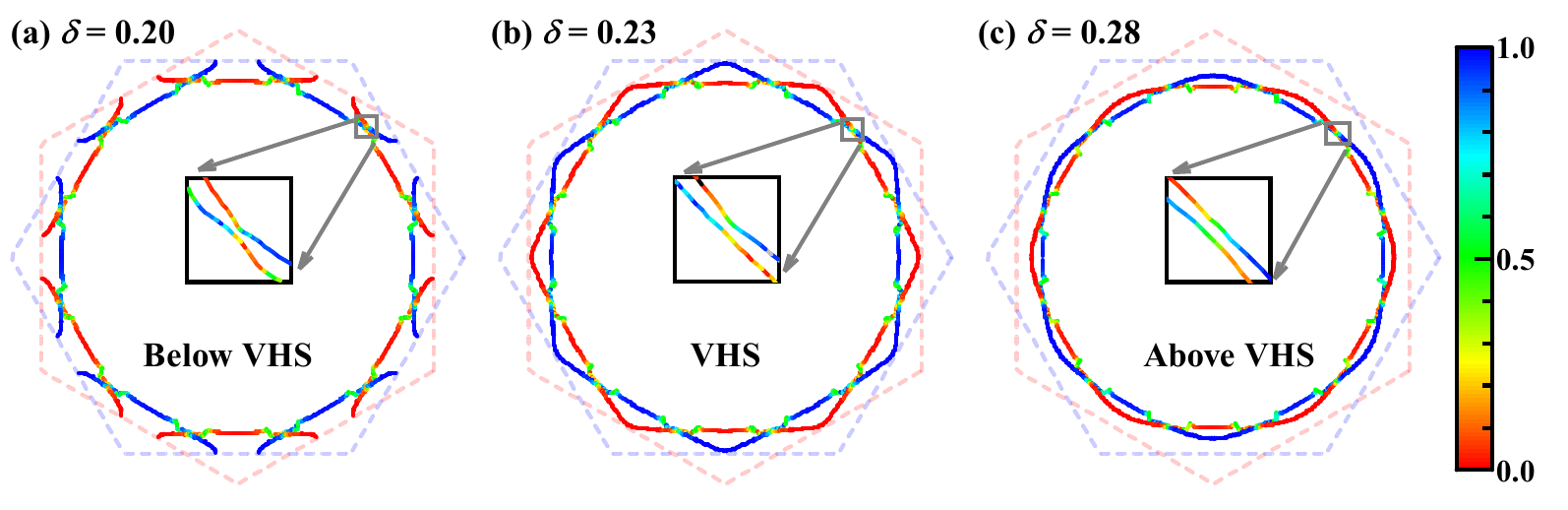}
	\caption{Fermi surfaces below (a), at (b), and above (c) the von Hove singularity (VHS) denoted in Fig.~\ref{dos-gap}(c). The blue and red dashed hexagons represent Brillouin zones of the top and bottom graphene layers, respectively. The color on the Fermi surfaces represents the occupation ratio of the top-graphene-layer bands.}
	\label{fs}
\end{figure}

In Fig.~\ref{fs}(a-c), we plot the FSs below, at, and above the VHS. As introduced on the above, due to the interlayer coupling, the inner and outer FSs formed from the hybridization of the FSs of the two mono-layers possess 12-folded symmetry, which is absent in a single monolayer FS. Note that the VH doping represents the doping level where the outer FS experiences the Lifshitz transition, which takes place at the $\mathbf{M}$ points on the outer FS, and changes the topology of the outer FS.

\section{Mean-field Theory }

This section provides some details of the mean-field calculations for the t-J model in the QC-TBG.

As a start point, the superexchange interaction part of the Hamiltonian can be separated into the spin singlet- and triplet-pairing channels, i.e.,
\begin{eqnarray}\label{MF}
H_J &=& \sum_{(\mathbf{i,j})}J_{\mathbf{ij}}\mathbf{S}_{\mathbf{i}}\cdot\mathbf{S}_{\mathbf{j}} \nonumber   \\
~&=& \sum_{(\mathbf{i,j})}J_{\mathbf{ij}}\left[\frac{1}{2}\left( c_{\mathbf{i}\uparrow}^{\dagger} c_{\mathbf{i}\downarrow} c_{\mathbf{j}\downarrow}^{\dagger} c_{\mathbf{j}\uparrow}+ c_{\mathbf{i}\downarrow}^{\dagger} c_{\mathbf{i}\uparrow} c_{\mathbf{j}\uparrow}^{\dagger} c_{\mathbf{j}\downarrow}\right)+\frac{1}{4}\left( c_{\mathbf{i}\uparrow}^{\dagger} c_{\mathbf{i}\uparrow}- c_{\mathbf{i}\downarrow}^{\dagger} c_{\mathbf{i}\downarrow}\right)\left( c_{\mathbf{j}\uparrow}^{\dagger} c_{\mathbf{j}\uparrow}- c_{\mathbf{j}\downarrow}^{\dagger} c_{\mathbf{j}\downarrow}\right)\right] \nonumber    \\
~&=&\sum_{(\mathbf{i,j})}J_{\mathbf{ij}}\left(-\frac{3}{4} \Delta_{\mathbf{ij}(0,0)}^{\dagger} \Delta_{\mathbf{ij}(0,0)}+\frac{1}{4} \Delta_{\mathbf{ij}(1,-1)}^{\dagger} \Delta_{\mathbf{ij}(1,-1)}+\frac{1}{4} \Delta_{\mathbf{ij}(1,0)}^{\dagger} \Delta_{\mathbf{ij}(1,0)}+\frac{1}{4} \Delta_{\mathbf{ij}(1,1)}^{\dagger} \Delta_{\mathbf{ij}(1,1)} \right),
\end{eqnarray}
where $J_{\mathbf{ij}}=4t_{\mathbf{ij}}^2/U$ and $U=10$ eV. The singlet-pairing channel $ \Delta_{\mathbf{ij}(0,0)}$ reads as
\begin{equation}
 \Delta_{\mathbf{ij}(0,0)}=\frac{1}{\sqrt{2}}\left( c_{\mathbf{i}\uparrow} c_{\mathbf{j}\downarrow}- c_{\mathbf{i}\downarrow} c_{\mathbf{j}\uparrow}\right)
\end{equation}
and the three degenerate components of the triplet pairing are as following,
\begin{eqnarray}
 \Delta_{\mathbf{ij}(1,1)}= c_{\mathbf{i}\uparrow} c_{\mathbf{j}\uparrow}, \quad
 \Delta_{\mathbf{ij}(1,0)}=\frac{1}{\sqrt{2}}\left( c_{\mathbf{i}\uparrow} c_{\mathbf{j}\downarrow}+ c_{\mathbf{i}\downarrow} c_{\mathbf{j}\uparrow}\right), \quad
 \Delta_{\mathbf{ij}(1,-1)}= c_{\mathbf{i}\downarrow} c_{\mathbf{j}\downarrow}.
\end{eqnarray}
Due to $J_{\mathbf{ij}}>0$, the triplet-pairing channels are always suppressed and thus only the singlet-pairing channel is considered, that is,
\begin{eqnarray}    \label{real_BCS}
H_J^{(s)} = -\frac{3}{4}\sum_{(\mathbf{i,j})}J_{\mathbf{ij}} \Delta_{\mathbf{ij}(0,0)}^{\dagger} \Delta_{\mathbf{ij}(0,0)}
=-\frac{3}{8}\sum_{(\mathbf{i,j})}J_{\mathbf{ij}}\left( c_{\mathbf{j}\downarrow}^{\dagger} c_{\mathbf{i}\uparrow}^{\dagger}- c_{\mathbf{j}\uparrow}^{\dagger} c_{\mathbf{i}\downarrow}^{\dagger}\right)\left( c_{\mathbf{i}\uparrow} c_{\mathbf{j}\downarrow}- c_{\mathbf{i}\downarrow} c_{\mathbf{j}\uparrow}\right).
\end{eqnarray}

This interaction Hamiltonian can be transformed to the eigen basis expanded by the perturbation-corrected eigenstates $\left\{|\widetilde{\mathbf{k}\alpha^{(\rm\mu)}}\rangle\right\}$. Our numerical results verify that each two states within this set are almost orthogonal to each other, which justifies this set as a good basis for the following study involving interaction. Writing the creation (annihilation) operator of the eigenstate $|\widetilde{\mathbf{k}\alpha^{(\rm\mu)}}\rangle$ for the spin $\sigma$ as $\tilde{c}^{\dagger}_{\mathbf{k}\mu\alpha\sigma}$ ($\tilde{c}_{\mathbf{k\mu}\alpha\sigma}$), we have
\begin{equation}\label{transform}
c_{\mathbf{i}\sigma}=\frac{1}{\sqrt{N}}\sum_{\mathbf{k\mu}\alpha}\tilde{c}_{\mathbf{k\mu}\alpha\sigma}\tilde\xi_{\mathbf{i},\mathbf{k\mu}\alpha}
\end{equation}
where $\tilde\xi_{\mathbf{i},\mathbf{k\mu}\alpha}/\sqrt{N}$ represents the real-space wave function for the perturbation-corrected eigenstate $|\widetilde{\mathbf{k}\alpha^{(\rm\mu)}}\rangle$. Substituting Eq. (\ref{transform}) back to Eq. (\ref{real_BCS}), we get the following BCS Hamiltonian,
\begin{eqnarray}    \label{k_BCS}
H_J^{(s)} &=&-\frac{3}{8N^2}\sum_{(\mathbf{i,j})}J_{\mathbf{ij}}\sum_{\mathbf{k}\mu\alpha\atop\mathbf{q}\nu\beta}
\tilde{c}_{\mathbf{k\mu\alpha}\downarrow}^{\dagger}
\tilde{c}_{\mathbf{-k\mu\alpha}\uparrow}^{\dagger}
\tilde{c}_{\mathbf{-q\nu\beta}\uparrow}
\tilde{c}_{\mathbf{q\nu\beta}\downarrow}
\left(\tilde\xi_{\mathbf{j},\mathbf{k}\mu\alpha}^{*}\tilde\xi_{\mathbf{i},-\mathbf{k}\mu\alpha}^{*}+\tilde\xi_{\mathbf{i},\mathbf{k}\mu\alpha}^{*}\tilde\xi_{\mathbf{j},-\mathbf{k}\mu\alpha}^{*}\right)
\left(\tilde\xi_{\mathbf{i},-\mathbf{q}\nu\beta}\tilde\xi_{\mathbf{j},\mathbf{q}\nu\beta}+\tilde\xi_{\mathbf{j},-\mathbf{q}\nu\beta}\tilde\xi_{\mathbf{i},\mathbf{q}\nu\beta}\right) \nonumber    \\
~&=&\sum_{\mathbf{k}\mu\alpha\atop\mathbf{q}\nu\beta}\frac{1}{N}
\tilde{c}_{\mathbf{k\mu\alpha}\downarrow}^{\dagger}
\tilde{c}_{\mathbf{-k\mu\alpha}\uparrow}^{\dagger}
\tilde{c}_{\mathbf{-q\nu\beta}\uparrow}
\tilde{c}_{\mathbf{q\nu\beta}\downarrow}
\left[-\frac{3}{2N}\sum_{(\mathbf{i,j})}J_{\mathbf{ij}}\mathbf{Re}(\tilde\xi_{\mathbf{i},\mathbf{k}\mu\alpha}\tilde\xi_{\mathbf{j},\mathbf{k}\mu\alpha}^{*})\mathbf{Re}(\tilde\xi_{\mathbf{i},\mathbf{q}\nu\beta}\tilde\xi_{\mathbf{j},\mathbf{q}\nu\beta}^{*})\right]\nonumber    \\
~&=&\sum_{\mathbf{k}\mu\alpha\atop\mathbf{q}\nu\beta}
\frac{1}{N}\tilde{c}_{\mathbf{k\mu\alpha}\downarrow}^{\dagger}
\tilde{c}_{\mathbf{-k\mu\alpha}\uparrow}^{\dagger}
\tilde{c}_{\mathbf{-q\nu\beta}\uparrow}
\tilde{c}_{\mathbf{q\nu\beta}\downarrow}
V^{\mu\nu}_{\alpha\beta}(\mathbf{k},\mathbf{q}),
\end{eqnarray}
with the pairing potential $V^{\mu\nu}_{\alpha\beta}(\mathbf{k},\mathbf{q})$ defined as
\begin{equation}
V^{\mu\nu}_{\alpha\beta}(\mathbf{k},\mathbf{q})=-\frac{3}{2N}\sum_{(\mathbf{i,j})}J_{\mathbf{ij}}\mathbf{Re}(\tilde\xi_{\mathbf{i},\mathbf{k}\mu\alpha}\tilde\xi_{\mathbf{j},\mathbf{k}\mu\alpha}^{*})\mathbf{Re}(\tilde\xi_{\mathbf{i},\mathbf{q}\nu\beta}\tilde\xi_{\mathbf{j},\mathbf{q}\nu\beta}^{*}).
\end{equation}
Here $\mathbf{Re}(z)$ indicates the real part of $z$. Note that we only consider the intra-band pairing with opposite momenta and spin here, i.e. the pairing between the time-reversal pair $|\widetilde{\mathbf{k}\alpha^{(\rm\mu)}}\uparrow\rangle$ and $|\widetilde{-\mathbf{k}\alpha^{(\rm\mu)}}\downarrow\rangle$.

The Hamiltonian in Eq.~\eqref{k_BCS} can be mean-field decoupled in the BCS channel to get the following mean-field Hamiltonian,
\begin{align}\label{HMF}
H_{\rm MF}=\sum_{\mathbf{k}\mu\alpha,\sigma}(\tilde{\varepsilon}^{\mu\alpha}_\mathbf{k}-\mu_c)
\tilde{c}^{\dagger}_{\mathbf{k}\mu\alpha\sigma}\tilde{c}_{\mathbf{k}\mu\alpha\sigma}
+\sum_{\mathbf{k}\mu\alpha}\left[
\Delta_{\mu\alpha}(\mathbf{k})
\tilde{c}^{\dagger}_{\mathbf{k}\mu\alpha\downarrow}
\tilde{c}^{\dagger}_{-\mathbf{k}\mu\alpha\uparrow}+h.c.\right]
\end{align}
where $\mu_c$ is the chemical potential. The superconductor pairing gap $\Delta_{\mu\alpha}(\mathbf{k})$ reads
\begin{align}\label{seq}
\Delta_{\mu\alpha}(\mathbf{k})=&\frac{1}{N}\sum_{\mathbf{q}\nu\beta}
V^{\mu\nu}_{\alpha\beta}(\mathbf{k},\mathbf{q})
\langle\tilde{c}_{-\mathbf{q}\nu\beta\uparrow}
\tilde{c}_{\mathbf{q}\nu\beta\downarrow}\rangle                                       \nonumber\\
=&-\frac{1}{N}\sum_{\mathbf{q}\nu\beta}V^{\mu\nu}_{\alpha\beta}(\mathbf{k},\mathbf{q})\frac{\Delta_{\nu\beta}(\mathbf{q})}
{2\sqrt{\left|\tilde{\varepsilon}^{\nu\beta}_\mathbf{q}-\mu_c\right|^2+\left|\Delta_{\nu\beta}(\mathbf{q})\right|^2}}
\left\{1-2f\left[\sqrt{\left|\tilde{\varepsilon}^{\nu\beta}_\mathbf{q}-\mu_c\right|^2
	+|\Delta_{\nu\beta}(\mathbf{q})|^2}\right]\right\}.
\end{align}
Here, $f[x]$ is the Fermi distribution function. Since $\Delta_{\mu\alpha}(\mathbf{k})\rightarrow 0$ when $T\rightarrow T_c$, Eq. \eqref{seq} leads to
\begin{align} \label{seq1}
\Delta_{\mu\alpha}(\mathbf{k})=&-\frac{1}{N}\sum_{\mathbf{q}\nu\beta}
V^{\mu\nu}_{\alpha\beta}(\mathbf{k},\mathbf{q})
\frac{\Delta_{\nu\beta}(\mathbf{q})}{2|\tilde{\varepsilon}^{\nu\beta}_\mathbf{q}-\mu_c|}
\tanh\left(\frac{|\tilde{\varepsilon}^{\nu\beta}_\mathbf{q}-\mu_c|}{2k_BT_c}\right)                   \nonumber\\
=&-\frac{1}{(2\pi)^2}\sum_{\nu\beta}\int d\mathbf{q}^2
V^{\mu\nu}_{\alpha\beta}(\mathbf{k},\mathbf{q})
\frac{\Delta_{\nu\beta}(\mathbf{q})}{2|\tilde{\varepsilon}^{\nu\beta}_\mathbf{q}-\mu_c|}
\tanh\left(\frac{|\tilde{\varepsilon}^{\nu\beta}_\mathbf{q}-\mu_c|}{2k_BT_c}\right).
\end{align}
After some further derivations \cite{4, 5}, we have
\begin{align}\label{linear_eq}
-\frac{1}{(2\pi)^2}\sum_{\nu\beta}\oint dq_{\parallel}
\frac{V^{\mu\nu}_{\alpha\beta}(\mathbf{k},\mathbf{q})}
{v^{\nu\beta}_F(\bm{q})}\Delta_{\nu\beta}(\mathbf{q})
=\lambda\Delta_{\mu\alpha}(\mathbf{k}),
\end{align}
where $v_F^{\nu\beta}(\mathbf{q})$ is the Fermi velocity and $q_{\parallel}$ denotes the component along the tangent of the FS. The pairing symmetry is determined by the gap form factor $\Delta_{\mu\alpha}(\mathbf{k})$ corresponding to the largest pairing eigenvalue $\lambda$ solved for this equation. The superconducting critical temperature $T_c$ is related to $\lambda$ via the relation $T_c\sim e^{-1/\lambda}$.

Note that on the above mean-field treatment, we have not considered the no-double-occupance constraint. This constraint can be treated by the Gutzwiller approximation, under which the free-electron part of the Hamiltonian will be renormalized by the Gutzwiller factor $\delta$. Under such treatment, the only revision of the above Eq.~\eqref{linear_eq} lies in that the Fermi velocity $v^{\nu\beta}_F(\bm{q})$ should be renormalized by the factor $\delta$. Under such revision of Eq.~\eqref{linear_eq}, the pairing symmetry would not be changed. Therefore, the no-double-occupance constraint would not change the pairing symmetry obtained by our mean-field calculations. The influence of this constraint on the superconducting $T_c$ is as follow. Firstly, as $v^{\nu\beta}_F(\bm{q})\to \delta v^{\nu\beta}_F(\bm{q})$, we have $\lambda\to \lambda/\delta$. Secondly, under the no-double-occupance constraint, the $T_c$ obtained via the relation $T_c\sim e^{-1/\lambda}$ only represents the temperature for the pseudo gap. As the Gutzwiller approximation renormalizes the pairing order parameter by the factor $\delta$, the real $T_c$ should also be renormalized by this factor. As a result, we have $T_c\sim \delta e^{-\delta /\lambda}$. For $\delta\to 0$, as $\lambda\propto \rho_{E_F}\propto \delta$, we have $T_c\propto \delta$. Here $\rho_{E_F}$ represents for the DOS on the FS. Near the VH doping, the $T_c$ estimated under our choice of interaction parameters, i.e. $U=10$ eV, is slightly lower than that obtained by the mean-field calculation. Overall, the $T_c\sim\delta$ curve after the Gutzwiller projection is qualitatively similar with that of the mean-field result. Note that as the accurate interaction strength of the real material cannot be exactly known, the strength of the superexchange coefficients $J_{\mathbf{ij}}$ can vary by several times from the adopted values in our study, which will considerably influence the obtained $T_c$. Physically, the superconducting $T_c$ of the QC-TBG should be near that of the monolayer graphene studied previously, as the pairing eigenvalue $\lambda$ obtained here is close to the value obtained for the case with the interlayer coupling turned off, as suggested by our numerical results.

\section{Pairing symmetry classification}

In the linearized gap equation \eqref{linear_eq}, as the pairing potential $V^{\mu\nu}_{\alpha\beta}(\mathbf{k},\mathbf{q})$ is invariant under the point-group symmetry of the system, all the obtained normalized gap function $\left\{\Delta_{\mu\alpha}(\mathbf{k})\right\}$ corresponding to the same eigenvalue $\lambda$ form an irreducible representation (IRRP) of the point group. Therefore, the pairing symmetries of the QC-TBG can be classified according to the IRRPs of the $D_{6d}$ point group. Redefining the 30$^\circ$-rotation about the centric axis (normal to the QC-TBG plane) as the combination of it and a succeeding layer exchange, the $D_{6d}$ point group can be redefined as the $D_{12}$ point group, whose IRRPs are shown in Table \ref{d12group}~\cite{x3}.

\begin{table*}[htbp]
	\caption{\label{tab:classification}IRRPs of the $D_{12}$ point group and classification of the pairing symmetries in the QC-TBG. The operator $\hat{P}_{\theta}$ denotes the rotation by the angle $\theta=n\pi/6$ ($n=0, 1,\cdots, 11$) about the centric axis (normal to the QC-TBG plane) and the operator $\hat{\sigma}$ reflects the QC-TBG about any of the twelve symmetric axes (in the QC-TBG plane). $D_{(\hat O)}$ is the matrix representation of $D_{12}$ group operators $\hat O$. $C_{\pi/6}$ and $\sigma_{x}$ are the two generators of $D_{12}$. For 2D IRRPs, $\Delta_{\mathbf{k}}=\Delta_{1\mathbf{k}}\pm i\Delta_{2\mathbf{k}}$ with $\Delta_{1\mathbf{k}}$ and $\Delta_{2\mathbf{k}}$ being the basis functions of the 2D IRRPs.}
	\begin{tabular}{|p{0.03\textwidth}<{\centering}p{0.03\textwidth}<{\centering}|p{0.18\textwidth}<{\centering}|p{0.07\textwidth}<{\centering}|p{0.28\textwidth}<{\centering}|p{0.27\textwidth}<{\centering}|}
		\hline
		\multicolumn{2}{|c|}{IRRPs}         &     $D_{(C_{\pi/6})}$                                      &   $D_{(\sigma_{x})}$                   &  Pairing symmetries            & Ground-state gap functions\\
		\hline
		\multirow{4}{*}{1D}      & \multicolumn{1}{|c|}{$A_1$} & $I$        &  $I$                      &  $s$
		& $\Delta_{\hat{P}_{\theta}\mathbf{k}}=\Delta_{\mathbf{k}}$, $\Delta_{\hat{\sigma} \mathbf{k}}=\Delta_{\mathbf{k}}$ \\
		\cline{2-6}
		& \multicolumn{1}{|c|}{$A_2$} & $-I$         &  $-I$                  &  $i'=i_{3(x^{5}y+xy^{5})+10x^{3}y^{3}}$
		& $\Delta_{\hat{P}_{\theta} \mathbf{k}}=-\Delta_{\mathbf{k}}$, $\Delta_{\hat{\sigma} \mathbf{k}}=-\Delta_{\mathbf{k}}$ \\
		\cline{2-6}
		& \multicolumn{1}{|c|}{$B_1$} & $-I$         &  $I$                  &  $i=i_{x^{6}-y^{6}+15(x^{2}y^{4}-x^{4}y^{2})}$
		& $\Delta_{\hat{P}_{\theta} \mathbf{k}}=-\Delta_{\mathbf{k}}$, $\Delta_{\hat{\sigma} \mathbf{k}}=\Delta_{\mathbf{k}}$ \\
		\cline{2-6}
		& \multicolumn{1}{|c|}{$B_2$} & $I$         &  $-I$                  &  $i*i'$
		& $\Delta_{\hat{P}_{\theta} \mathbf{k}}=\Delta_{\mathbf{k}}$, $\Delta_{\hat{\sigma} \mathbf{k}}=-\Delta_{\mathbf{k}}$ \\
		\hline
		\multirow{5}{*}{2D}      & \multicolumn{1}{|c|}{$E_1$} & $I\cos\frac{\pi}{6} - i\sigma_{y}\sin\frac{\pi}{6}$                      &  $-\sigma_{z}$ & $(p_{x}, p_{y})$
		& $\Delta_{\hat{P}_{\theta}\mathbf{k}}=e^{\pm i{\pi\over6}}\Delta_{\mathbf{k}}$, $\Delta_{\hat{\sigma} \mathbf{k}} \neq\Delta_{\mathbf{k}}$ \\
		\cline{2-6}
		& \multicolumn{1}{|c|}{$E_2$} &  $I\cos\frac{\pi}{3} - i\sigma_{y}\sin\frac{\pi}{3}$    &  $\sigma_{z}$               &  $(d_{x^2-y^2},d_{xy})$
		&$\Delta_{\hat{P}_{\theta}\mathbf{k}}=e^{\pm i{\pi\over3}}\Delta_{\mathbf{k}}$, $\Delta_{\hat{\sigma} \mathbf{k}} \neq\Delta_{\mathbf{k}}$ \\
		\cline{2-6}
		& \multicolumn{1}{|c|}{$E_3$} &  $-i\sigma_{y}$    &  $-\sigma_{z}$               &  $(f_{x^3-3xy^2},f_{3x^2y-y^3})$
		&$\Delta_{\hat{P}_{\theta}\mathbf{k}}=e^{\pm i{\pi\over2}}\Delta_{\mathbf{k}}$, $\Delta_{\hat{\sigma} \mathbf{k}} \neq\Delta_{\mathbf{k}}$ \\
		\cline{2-6}
		& \multicolumn{1}{|c|}{$E_4$} &  $I\cos\frac{2\pi}{3} - i\sigma_{y}\sin\frac{2\pi}{3}$    &  $\sigma_{z}$               &  $(g_{x^4+y^4-6x^2y^2},g_{x^3y-xy^3})$
		&$\Delta_{\hat{P}_{\theta}\mathbf{k}}=e^{\pm i{2\pi\over3}}\Delta_{\mathbf{k}}$, $\Delta_{\hat{\sigma} \mathbf{k}} \neq\Delta_{\mathbf{k}}$ \\
		\cline{2-6}
		& \multicolumn{1}{|c|}{$E_5$} &  $I\cos\frac{5\pi}{6} - i\sigma_{y}\sin\frac{5\pi}{6}$    &  $-\sigma_{z}$               &  $(h_{x^5+5xy^4-10x^3y^2},h_{y^5+5x^4y-10x^2y^3})$
		&$\Delta_{\hat{P}_{\theta}\mathbf{k}}=e^{\pm i{5\pi\over6}}\Delta_{\mathbf{k}}$, $\Delta_{\hat{\sigma} \mathbf{k}} \neq\Delta_{\mathbf{k}}$ \\
		\hline
	\end{tabular}
	\label{d12group}
\end{table*}

In Table \ref{d12group}, the second and third columns list the representation matrices of the two generators of the $D_{12}$ point group up to a global unitary transformation. In the fourth column, we list all the possible pairing symmetries, in which the 1D IRRPs  includes the $s$-wave with angular momentum $L=0$, the $i=i_{x^{6}-y^{6}+15(x^{2}y^{4}-x^{4}y^{2})}$ and $i'=i_{3(x^{5}y+xy^{5})+10x^{3}y^{3}}$ waves with $L=6$, and the $i*i'$ wave with $L=0$, and the 2D IRRPs include the $(p_x,p_y)$ wave with $L=1$, the $(d_{x^2-y^2},d_{xy})$ wave with $L=2$, the $(f_{x^3-3xy^2},f_{3x^2y-y^3})$ wave with $L=3$, the $(g_{x^4+y^4-6x^2y^2},g_{x^3y-xy^3})$ wave with $L=4$, and the $(h_{x^5+5xy^4-10x^3y^2},h_{y^5+5x^4y-10x^2y^3})$ wave with $L=5$. Note that as we only consider the intra-band pairing between opposite momenta, the spin statistics and the pairing angular momentum are mutually determined: for the spin-singlet pairings, $L$ should be even, while for the spin-triplet pairings, $L$ should be odd. This point is in contrast with the intrinsic QC wherein the spin statistics and the pairing angular momentum are independent \cite{x1}.  In the t-J model studied here, only the singlet pairings with even $L$ are possible, including the $s$-, $(d_{x^2-y^2},d_{xy})$-, $(g_{x^4+y^4-6x^2y^2},g_{x^3y-xy^3})$-, $i_{x^{6}-y^{6}+15(x^{2}y^{4}-x^{4}y^{2})}$-, $i_{3(x^{5}y+xy^{5})+10x^{3}y^{3}}$-, and the $i*i'$- wave pairing symmetries. However, in other model, such as the Hubbard model, all the other triplet pairings with odd $L$ are also possible, which are not studied in the present work.

In the last column of Table \ref{d12group}, we show the properties of the ground-state wave function of each pairing symmetry. For the 1D IRRPs, the symmetries of the ground-state wave functions are the same as those obtained from solving Eq.~\eqref{linear_eq} at $T_c$. For a 2D IRRP, the ground-state wave function is the mixing of the two basis functions of that IRRP. From the Ginzburg-Landau theory provided in the next section in combination with our numerical calculations for the cases of degenerate $d_{x^2-y^2}$ and $d_{xy}$ waves or degenerate $g_{x^4+y^4-6x^2y^2}$ and $g_{x^3y-xy^3}$ waves, the two basis functions would always be mixed as $1:\pm i$.  Under such a mixing manner, the ground-state wave functions are complex, whose complex phases change $\pm L\theta$ with each rotation by the angle $\theta$, where $\theta=n\pi/6$. In the meantime, the mirror reflection symmetry $\sigma$ is broken in these 2D IRRPs ground states. For the other cases of 2D IRRPs, including the $(p_x,p_y)$, the $(f_{x^3-3xy^2},f_{3x^2y-y^3})$ and the $(h_{x^5+5xy^4-10x^3y^2},h_{y^5+5x^4y-10x^2y^3})$ which can emerge in other models, the situations are similar. The Ginzburg-Landau theory for arguing the $1:\pm i$ mixing for these cases also applies. Although no numerical calculation is available for these cases, the general physical reason for this mixing is the same: the $1:\pm i$ mixing manner can lead to a full pairing gap to minimize the free energy, while other solutions for the minimum of the Ginzburg-Landau free energy cannot.

The main features exhibited in Table \ref{d12group} also apply to other 2D lattices with $D_{2n}$ point group, including periodic lattices with $n\le3$. The $D_{2n}$ point group possesses $4n$ elements, which are divided into $(n+3)$ classes. It possesses four 1D IRRPs and $(n-1)$ 2D IRRPs, satisfying $4+(n-1)=n+3$ and $4\times1^2+(n-1)\times2^2=4n$.  The four 1D IRRPs include the $A_1,A_2,B_1$, and $B_2$ representations. The $A_1$ representation is the identity one, corresponding to the $s$-wave. The $B_1$ and $A_2$ representations correspond to the pairing states with the largest angular momentum $L=n$, similar to the $i$- and $i'$-wave pairings here. The $B_2$ representation is the product representation of $B_1$ and $A_2$. The $(n-1)$ 2D IRRPs are marked as $E_L$, with $L=1,\cdots,(n-1)$. The two lowest-harmonics basis functions of the $E_L$ IRRP can be chosen as the real and imaginary parts of $(x+yi)^L$, respectively. When they are $1:\pm i$ mixed, the resulting complex gap function behaves as such: with every $\pi/n$ rotation, the phase angle of the complex pairing gap function changes $\pm L\pi/n$. Therefore, the index $L$ is just the pairing angular momentum of the topological superconductivity (TSC) on a 2D lattice hosting the $D_{2n}$ point group, which is at most $(n-1)$. On periodic lattices, since $n\le 3$, one can only get $p+ip$ TSC with $L=1$ or $d+id$ TSC with $L=2$. For example, on square lattices with $D_4$ point group, since $n=2$, the largest pairing angular momentum of the TSCs on square lattice is $L=n-1=1$, which only includes the $p+ip$ TSC. On the honeycomb lattices with $D_6$ point group, since $n=3$, the largest $L=n-1=2$. Therefore, the TSCs on the honeycomb lattices can be either $p+ip$ ($L=1$) or $d+id$ ($L=2$). Occasionally, the gap functions belonging to the $E_L$ IRRP can take higher-harmonics formulism such as $(x+yi)^{L\pm n}$, $(x+yi)^{L\pm 2n}$, ....... In such cases, on, say the honeycomb lattice, the $g+ig$ can also emerge as the higher-harmonics component of the $E_2$ IRRP~\cite{8}. However, since the $g+ig$ and the $d+id$ on the honeycomb lattice both belong to $E_2$, they would generally be mixed. For a pairing gap function with mixed $d+id$ and $g+ig$ components belonging to the same $E_2$ IRRP of $D_6$, one usually attributes its pairing symmetry to the lowest-harmonics component, i.e. $d+id$, unless the weight of that component is zero or extremely low~\cite{8}. Such situation only happens occasionally since no symmetry can guarantee it. Therefore, on the honeycomb lattice, we can only have accidental $g+ig$ TSC for the $E_2$ IRRP, or accidental $h+ih$ TSC for the $E_1$ IRRP. Similarly, on the square lattice, we can only have accidental $f+if$ TSC for the $E_1$ IRRP. However, on the QC-TBG with effective $D_{12}$ point-group symmetry, the $d+id$ and $g+ig$ belong to $E_2$ and $E_4$ IRRPs respectively, which by symmetry will not be mixed. Therefore, the $g+ig$ obtained in the QC-TBG is protected by symmetry, instead of being accidental.

\begin{figure}[htbp]
	\centering
	\includegraphics[width=0.95\textwidth]{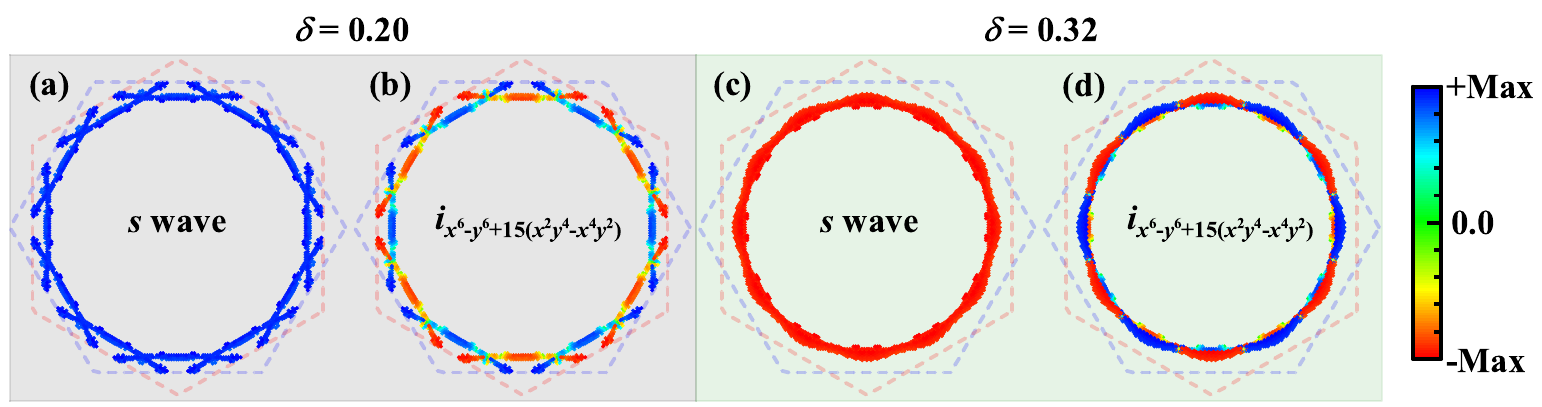}
	\caption{Distributions of paring gap functions for $s$- and $i$-waves on the Fermi surfaces when $\delta=0.20$ and $0.32$. }
	\label{si}
\end{figure}

\begin{figure}[htbp]
	\centering
	\includegraphics[width=0.9\textwidth]{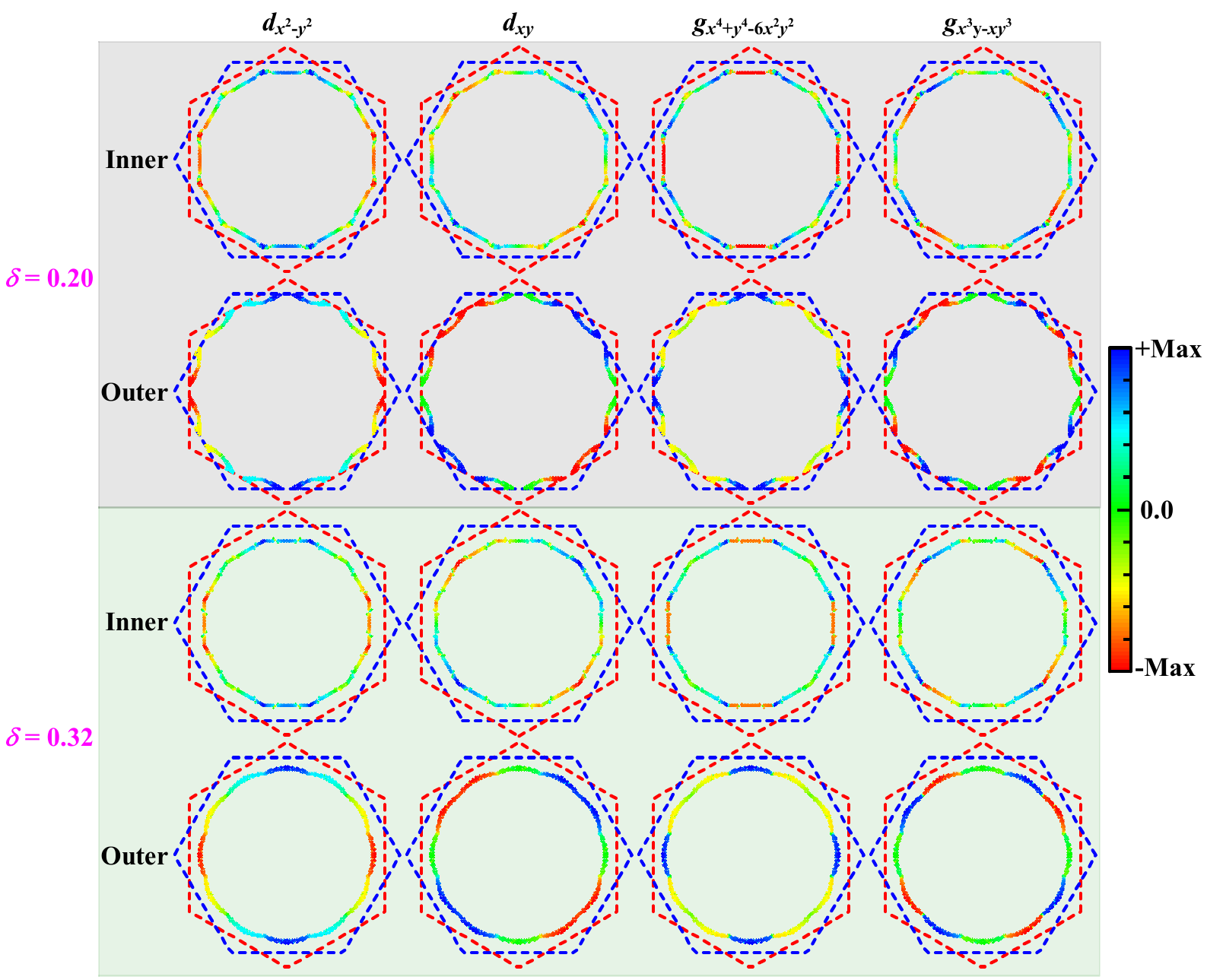}
	\caption{Distributions of paring gap functions for $d$- and $g$-waves on the inner and outer Fermi surfaces when $\delta=0.20$ and $0.32$.}
	\label{dg}
\end{figure}

\section{  More informations on gap functions}

This section provides more informations about the four leading gap functions obtained from solving the linearized gap equation \eqref{linear_eq}. Here we will show the gap functions of the $s$-, the $i_{x^{6}-y^{6}+15(x^{2}y^{4}-x^{4}y^{2})}$-, the $(d_{x^2-y^2},d_{xy})$, and the $(g_{x^4+y^4-6x^2y^2},g_{x^3y-xy^3})$ wave pairings for two typical doping levels, i.e. $\delta=0.20$ below the VH doping and $\delta=0.32$ above the VH doping. The other two pairing symmetries, i.e. the $i_{3(x^{5}y+xy^{5})+10x^{3}y^{3}}$ and $i*i'$ are not shown as their pairing eigenvalues are lower than the four shown symmetries. For the degenerate $(d_{x^2-y^2},d_{xy})$- and degenerate $(g_{x^4+y^4-6x^2y^2},g_{x^3y-xy^3})$- wave pairings, we shall determine how the two basis functions are mixed to minimize the ground-state energy by numerical calculations.

Figure~\ref{si} shows the leading gap functions of the $s$- and $i_{x^{6}-y^{6}+15(x^{2}y^{4}-x^{4}y^{2})}$-wave pairings corresponding to the largest pairing eigenvalues of the two pairing symmetries under $\delta=0.20$ and 0.32. Obviously, the $s$-wave pairing gap function is invariant under the $D_{12}$ point group. The $i$-wave pairing gap function changes sign for every $30\degree$ rotation. The gap function of this pairing symmetry is reflection even about the axes with azimuthal angles $\theta=n\pi/6$ ($n=-5,\cdots,6$) and reflection odd about the axes with azimuthal angles $\theta=(2n-1)\pi/12$ ($n=-5,\cdots,6$). The nodal directions of the pairing gap function for this pairing symmetry are at $\theta=(2n-1)\pi/12$ ($n=-5,\cdots,6$). These two pairing symmetries are consistent with the $A_1$ and $B_1$ IRRPs listed in Table \ref{d12group}, respectively.

Figure~\ref{dg} shows the leading gap functions of the degenerate $(d_{x^2-y^2},d_{xy})$- and $(g_{x^4+y^4-6x^2y^2},g_{x^3y-xy^3})$- wave pairing symmetries corresponding to their largest pairing eigenvalues under $\delta=0.20$ and 0.32. As the two pockets are close in the Brillioun zone, we plot the distributions of the gap functions on the inner and outer pockets separately to enhance the visibility. Figure~\ref{dg} informs us the following characters of these gap functions. Firstly, while the $d_{x^2-y^2}$- and the $g_{x^4+y^4-6x^2y^2}$- wave pairing gap functions are reflection even about the $x$- and $y$- axes, the $d_{xy}$- and the $g_{x^3y-xy^3}$- wave ones are reflection odd about these axes. Secondly, while the two $d$-wave pairing gap functions change sign for every 90\degree rotation, the two $g$-wave ones keep unchanged for such rotation. Thirdly, while each $d$-wave pairing gap function possesses four nodal points on each pocket, each $g$-wave pairing gap function possesses eight nodal points on each pocket. Finally, the nodal points for the two $d$-wave gap functions don't coincide with each other, and neither do those for the two $g$- wave ones.

Since the two $d$- and $g$- wave pairing gap functions each are doubly degenerate, we shall mix the two basis functions for each case to minimize the ground-state energy.  The start point is the Hamiltonian of the system,
\begin{align}\label{Hamiltonian}
	H=\sum_{\mathbf{k}\mu\alpha,\sigma}\tilde{\varepsilon}^{\mu\alpha}_\mathbf{k}
	\tilde{c}^{\dagger}_{\mathbf{k}\mu\alpha\sigma}\tilde{c}_{\mathbf{k}\mu\alpha\sigma}
	+\frac{1}{N}\sum_{\mathbf{k}\mu\alpha\atop\mathbf{q}\nu\beta}V^{\mu\nu}_{\alpha\beta}(\mathbf{k},\mathbf{q})
	\tilde{c}^{\dagger}_{\mathbf{k}\mu\alpha\downarrow}\tilde{c}^{\dagger}_{-\mathbf{k}\mu\alpha\uparrow}
	\tilde{c}_{-\mathbf{q}\nu\beta\uparrow}\tilde{c}_{\mathbf{q}\nu\beta\downarrow}.
\end{align}
We introduce the following pairing gap function,
\begin{align}
	\Delta_{\mu\alpha}(\mathbf{k})=\psi_{1}\Delta^{(1)}_{\mu\alpha}(\mathbf{k})
	+\psi_{2}\Delta^{(2)}_{\mu\alpha}(\mathbf{k}).
	\label{cpf}
\end{align}
where $\psi_{1}$ and $\psi_{2}$ are two complex numbers. The $\Delta^{(1)}_{\mu\alpha}(\mathbf{k})$ and $\Delta^{(2)}_{\mu\alpha}(\mathbf{k})$ represent the two degenerate normalized basis functions of the $d$- wave or $g$- wave pairing symmetries obtained from solving the linearized gap equation (\ref{linear_eq}). Using Eq.~\eqref{cpf}, we obtain the following mean-field Hamiltonian,
\begin{align}
	H_{\rm MF}=\sum_{\mathbf{k}\mu\alpha,\sigma}\left(\tilde{\varepsilon}^{\mu\alpha}_\mathbf{k}-\mu_c\right)
	\tilde{c}^{\dagger}_{\mathbf{k}\mu\alpha\sigma}\tilde{c}_{\mathbf{k}\mu\alpha\sigma}
	+\sum_{\mathbf{k}\mu\alpha}\left(\Delta_{\mu\alpha}(\mathbf{k})
	\tilde{c}^{\dagger}_{\mathbf{k}\mu\alpha\downarrow}
	\tilde{c}^{\dagger}_{-\mathbf{k}\mu\alpha\uparrow}+h.c.\right)
\end{align}
This Hamiltonian can be diagonalized to obtain the mean-field BCS ground state. Then $\psi_{1}$ and $\psi_{2}$ are determined by minimizing the ground-state energy, i.e. the expectation value $E$ of the Hamiltonian Eq.~\eqref{Hamiltonian} in the mean-field BCS ground state,
\begin{align}\label{Etot}
E=&\sum_{\mathbf{k}\mu\alpha,\sigma}\tilde{\varepsilon}^{\mu\alpha}_\mathbf{k}
	\left\langle\tilde{c}^{\dagger}_{\mathbf{k}\mu\alpha\sigma}\tilde{c}_{\mathbf{k}\mu\alpha\sigma}\right\rangle
	+\frac{1}{N}\sum_{\mathbf{k}\mu\alpha\atop\mathbf{q}\nu\beta}V^{\mu\nu}_{\alpha\beta}(\mathbf{k},\mathbf{q})
	\left\langle\tilde{c}^{\dagger}_{\mathbf{k}\mu\alpha\downarrow}\tilde{c}^{\dagger}_{-\mathbf{k}\mu\alpha\uparrow}\right\rangle
	\left\langle\tilde{c}_{-\mathbf{q}\nu\beta\uparrow}\tilde{c}_{\mathbf{q}\nu\beta\downarrow}\right\rangle
 \nonumber \\
=&\sum_{\mathbf{k}\mu\alpha}\tilde{\varepsilon}^{\mu\alpha}_\mathbf{k}
\left[1-\frac{\tilde{\varepsilon}^{\mu\alpha}_\mathbf{k}-\mu_c}
{\sqrt{(\tilde{\varepsilon}^{\mu\alpha}_\mathbf{k}-\mu_c)^2+
		|\Delta_{\mu\alpha}(\mathbf{k})|^2}}\right] \nonumber\\
&+\frac{1}{4N}\sum_{\mathbf{k}\mu\alpha\atop\mathbf{q}\nu\beta}V^{\mu\nu}_{\alpha\beta}(\mathbf{k},\mathbf{q})
\frac{\Delta^*_{\mu\alpha}(\mathbf{k})}
{\sqrt{(\tilde{\varepsilon}^{\mu\alpha}_\mathbf{k}-\mu_c)^2+
		|\Delta_{\mu\alpha}(\mathbf{k})|^2}}
\frac{\Delta_{\nu\beta}(\mathbf{q})}
{\sqrt{(\tilde{\varepsilon}^{\nu\beta}_\mathbf{q}-\mu_c)^2+
		|\Delta_{\nu\beta}(\mathbf{q})|^2}}.
\end{align}
Note that the chemical potential $\mu_c$ is determined by fixing the total electron number when varying $\psi_1$ and $\psi_2$.

\begin{figure}[htbp]
	\centering
	\includegraphics[width=0.5\textwidth]{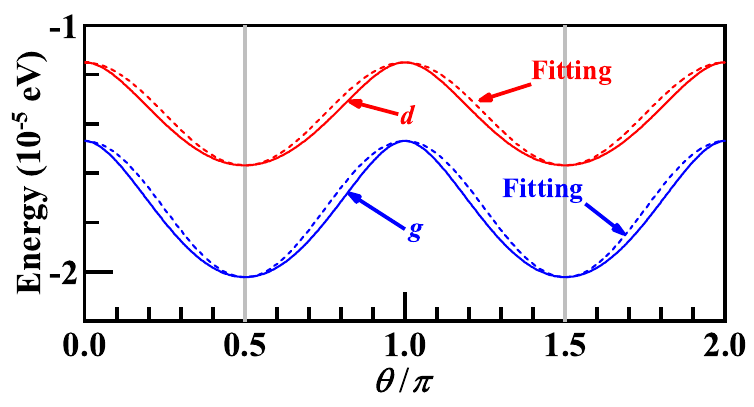}
	\caption{Variations of the energies $E$ with the mixing-phase-angle $\theta$ for the degenerate $d$- and $g$-wave pairings, with their global pairing amplitudes optimized for the energy minimization. The dashed lines represent the fittings of the cosine functions with the formula $E\left(\theta\right)=E_0+\eta\cos 2\theta$, with $\eta>0$.}
	\label{fit}
\end{figure}

Our numerical results for the energy minimization are as follow. Setting $\psi_1:\psi_2=1:\alpha e^{i\theta}$, our results suggest $\alpha=1$ and $E$ as functions of $\theta$, i.e. $E(\theta)$ are shown in Fig. \ref{fit} for both the $d$- and $g$- wave pairings. We can verify that both $E\sim \theta$ relation curve can be well fitted by the dashed lines described by the relations
\begin{equation}\label{energy_function}
E\left(\theta\right)=E_0+\eta\cos 2\theta
\end{equation}
with $\eta>0$. Consequently, the minimized energy is realized at $\theta=\pi/2$ or $3\pi/2$, leading to $\psi_1:\psi_2=1:\pm i$. Therefore, the ground-state pairing gap functions of the two pairing symmetries take the form of $d_{x^2-y^2}\pm i d_{xy}$- and $g_{x^4+y^4-6x^2y^2}\pm i g_{x^3y-xy^3}$, abbreviated as $d+id$ and $g+ig$. For this result, we will provide an understanding based on the Ginzburg-Landau theory in the next section.

\section{The Ginzburg-Landau theory}

In this section, we provide two Ginzburg-Landau (G-L) theory based analysises toward the above numerical results obtained by our MF calculations. In the first G-L theory, we provide an understanding for why the two basis functions of the obtained degenerate $(d_{x^2-y^2}, i d_{xy})$- or $(g_{x^4+y^4-6x^2y^2}, i g_{x^3y-xy^3})$- wave pairings in the QC-TBG should be mixed as $1:\pm i$. In the second one, we provide an understanding for why the coupling between the $d+id$ pairing order parameters in the two mono-layers can lead to either $d+id$ or $g+ig$ TSCs in the QC-TBG.

The first Ginzburg-Landau theory based analysis is as follow. We rewrite the Eq. (\ref{cpf}) into the following form as
\begin{eqnarray}
\Delta_{\mu\alpha}(\mathbf{k}) = \left(\begin{array}{cc}
\psi_{1} & \psi_{2}\end{array}\right)\left(\begin{array}{c}
\Delta_{\mu\alpha}^{(1)}(\mathbf{k})\\
\Delta_{\mu\alpha}^{(2)}(\mathbf{k})
\end{array}\right) \equiv {\bf\Psi}^T\left(\begin{array}{c}
\Delta_{\mu\alpha}^{(1)}(\mathbf{k})\\
\Delta_{\mu\alpha}^{(2)}(\mathbf{k})
\end{array}\right),
\end{eqnarray}
with $\bf {\Psi}\equiv (\psi_1,\psi_2)^T$. The free energy function $F$ can be expanded up to O($|\psi|^{4}$) as follow,
\begin{eqnarray}    \label{freeE}
F({\bf\Psi}) &=&
C_1\left(|\psi_{1}|^2+|\psi_{2}|^2\right)+C_2\left(|\psi_{1}|^2-|\psi_{2}|^2\right)+
C_3\left(\psi_{1}^{*}\psi_{2}+\psi_{1}\psi_{2}^{*}\right)+
C_4\left|\psi_{1}^{2}+\psi_{2}^{2}\right|^{2}+
C_5\left|\psi_{1}^{2}-\psi_{2}^{2}\right|^{2}\nonumber \\& &+
C_6(\psi_{1}^{*}\psi_{2}+\psi_{1}\psi_{2}^{*})\left(|\psi_{1}|^{2}+|\psi_{2}|^{2}\right)+C_7(\psi_{1}^{*}\psi_{2}+\psi_{1}\psi_{2}^{*})\left(|\psi_{1}|^{2}-|\psi_{2}|^{2}\right) \nonumber\\
& &+C_8\left(\left|\psi_{1}\right|^{2}+\left|\psi_{2}\right|^{2}\right)^2+C_9\left(\left|\psi_{1}\right|^{2}-\left|\psi_{2}\right|^{2}\right)^2+O(|\psi|^{6})  \nonumber\\
&=&
C_1\mathbf{\Psi}^\dagger \mathbf{\Psi}+C_2\mathbf{\Psi}^\dagger\sigma_{z}\mathbf{\Psi}+
C_3\mathbf{\Psi}^\dagger\sigma_{x}\mathbf{\Psi}+
C_4\left|\mathbf{\Psi}^T\mathbf{\Psi}\right|^{2}+
C_5\left|\mathbf{\Psi}^T\sigma_z\mathbf{\Psi}\right|^{2}+
C_6\left(\mathbf{\Psi}^\dagger\sigma_{x}\mathbf{\Psi}\right)\left(\mathbf{\Psi}^\dag\mathbf{\Psi}\right)\nonumber\\& &+
C_7\left(\mathbf{\Psi}^\dagger\sigma_{x}\mathbf{\Psi}\right)\left(\mathbf{\Psi}^\dag\sigma_z\mathbf{\Psi}\right)+
C_8\left(\mathbf{\Psi}^{\dagger}\mathbf{\Psi}\right)^{2}+
C_9\left(\mathbf{\Psi}^{\dagger}\sigma_z\mathbf{\Psi}\right)^{2}+O(|\psi|^{6}),
\end{eqnarray}
with $C_i\ (i=1,2,\cdots,9)\in R$. In obtaining this formula, we have used the U(1)-gauge and the time-reversal symmetries of the free energy which requires invariance of $F$ under the transformation $\mathbf{\Psi}\to e^{i\theta}\mathbf{\Psi}$ and and the one $\mathbf{\Psi}\to \mathbf{\Psi}^*$.

Applying an arbitrary operator $\hat{P}$ within the point group $D_{12}$ to the pairing gap function, we get
\begin{eqnarray}
\hat{P}\Delta_{\mu\alpha}(\mathbf{k})= {\bf\Psi}^T\hat{P}\left(\begin{array}{c}
\Delta_{\mu\alpha}^{(1)}(\mathbf{k})\\
\Delta_{\mu\alpha}^{(2)}(\mathbf{k})
\end{array}\right)= {\bf\Psi}^T\left(\begin{array}{c}
\Delta_{\mu\alpha}^{(1)'}(\mathbf{k})\\
\Delta_{\mu\alpha}^{(2)'}(\mathbf{k})
\end{array}\right)\equiv \tilde{\bf\Psi}^T\left(\begin{array}{c}
\Delta_{\mu\alpha}^{(1)}(\mathbf{k})\\
\Delta_{\mu\alpha}^{(2)}(\mathbf{k})
\end{array}\right),
\end{eqnarray}
with $\tilde{\bf\Psi}\equiv P^T\bf\Psi$, where $P$ is the matrix representation of $\hat{P}$. For the pure rotation part of $D_{12}$ with angle $\theta$, we have $P^T(\theta)=\cos \theta+i\sigma_y\sin \theta$.  Since the symmetry operator $\hat{P}$ does not change the free energy function, i.e., $F(\mathbf{\Psi})=F(\tilde{\mathbf{\Psi}})$, one can immediately obtain $C_2 = C_3=C_5 = C_6 = C_7=C_9=0$ because $\sigma_{x}$ and $\sigma_z$ in these terms do not commute with $\sigma_y$. As a result, the free energy function can be simplified as,
\begin{eqnarray}
F(\mathbf{\Psi}) = C_1\left(|\psi_{1}|^2+|\psi_{2}|^2\right)+
C_4\left|\psi_{1}^{2}+\psi_{2}^{2}\right|^{2}+
C_8\left(\left|\psi_{1}\right|^{2}+\left|\psi_{2}\right|^{2}\right)^2 + O(|\psi|^{6}).
\end{eqnarray}
Notice that this formula is consistent with that obtained in Ref~\cite{x2} for the honeycomb lattice with $D_6$ point-group. We neglect the $O(|\psi|^{6})$ term and assume the minimized free energy is realized at $\psi_{2}=e^{i\theta}\psi_{1}$, then
\begin{eqnarray}
F(\mathbf{\Psi}) = 2C_1|\psi_1|^2 + 2C_4|\psi_{1}|^{4} [\cos (2\theta)+1] + 4C_8|\psi_1|^4.
\end{eqnarray}
When $C_4>0$ the minimization of $F$ requires $\theta={\pi\over 2}$ or ${3\pi\over2}$, that is, $\psi_{1}:\psi_{2}=1: (\pm i)$. This is why our calculation results can well be fitted with the cosine function, as shown in Fig.~\ref{fit}. When $C_4<0$ the minimization of $F$ requires $\theta=0$ or $\pi$, that is, $\psi_{1}:\psi_{2}=1: (\pm 1)$. Although from the G-L theory alone, one doesn't know the sign of $C_4$, physically the solution $\psi_{1}:\psi_{2}=1: (\pm i)$ is more reasonable because in such cases the obtained pairing gap function is fully gapped, which benefits the energy gain. This Ginzburg-Landau theory based argument also applies to other degenerate pairing symmetries listed in the Table. \ref{tab:classification}, which can emerge in other models such as the Hubbard model on the QC-TBG.

The second G-L theory based analysis is as follow. In the QC-TBG, there are two graphene mono-layers which are weakly coupled. It's known that the electron-electron interaction within each mono-layer can induce degenerate $(d_{xy},d_{x^2-y^2})$- wave pairing. Therefore, if we turn off the inter-layer coupling, we would obtain four degenerate $d$-wave pairings, i.e. $(d^{(\text{t})}_{xy},d^{(\text{t})}_{x^2-y^2},d^{(\text{b})}_{xy},d^{(\text{b})}_{x^2-y^2})$. As the intra-layer hopping and interaction strengths are much stronger than the inter-layer ones, one can reasonably first let each two degenerate $d$-wave pairings on one mono-layer mix as $d\pm id$, and then consider their inter-layer Josephson coupling. Further more, we only consider the cases wherein the pairings from the two mono-layers are both $d+id$ or $d-id$, because otherwise the system cannot gain energy from the inter-layer Josephson coupling. The argument for this point is as follow.

Physically, the coupling between the $d\pm id$ pairing order parameters from the two layers mainly originates from the inter-layer Josephson coupling. Formally, this coupling can be understood as the consequence of the second-order perturbational process, if one takes the inter-layer tunneling term $H'$ in Eq. (\ref{perturbation}) as the perturbation. The contribution of this second-order perturbation to the energy or the free energy can be roughly taken as
\begin{eqnarray}\label{Josephson0}
F_{inter-layer}\approx -\frac{\left\langle H'^2\right\rangle}{2\Delta},H'= -\sum_{\textbf{i}\textbf{j}\sigma}  c_{\textbf{i}\rm{t}\sigma}^{\dagger} c_{\textbf{j}\rm{b}\sigma}t_{\textbf{i}\textbf{j}}+h.c.,
\end{eqnarray}
where $\Delta$ denotes the averaged pairing-gap amplitude and $t_{\textbf{i}\textbf{j}}$ is symmetric under the 6-folded rotation. From Eq. (\ref{Josephson0}), we have
\begin{eqnarray}\label{Josephson1}
F_{inter-layer}&\approx&  -\frac{1}{2\Delta}\sum_{\mathbf{i}\mathbf{j}\sigma \atop \tilde{\mathbf{i}}\tilde{\mathbf{j}}\tilde{\sigma}}\left\langle \left(c_{\textbf{i}\rm{t}\sigma}^{\dag} c_{\textbf{j}\rm{b}\sigma}t_{\textbf{i}\textbf{j}}+h.c.\right)\left(c_{\tilde{\textbf{i}}\rm{t}\tilde{\sigma}}^{\dag} c_{\tilde{\textbf{j}}\rm{b}\tilde{\sigma}}t_{\tilde{\textbf{i}}\tilde{\textbf{j}}}+h.c.\right)\right\rangle\nonumber\\
&\approx& -\frac{1}{2\Delta}\sum_{\mathbf{i}\mathbf{j}\tilde{\mathbf{i}}\tilde{\mathbf{j}}\sigma}\left\langle c_{\textbf{i}\rm{t}\sigma}^{\dag}c_{\tilde{\textbf{i}}\rm {t}\bar{\sigma}}^{\dag}\right\rangle \left\langle c_{\tilde{\textbf{j}}\rm {b}\bar{\sigma}}c_{\textbf{j}\rm {b}\sigma}\right\rangle t_{\textbf{i}\textbf{j}} t_{\tilde{\textbf{i}}\tilde{\textbf{j}}}+c.c. = -\frac{1}{2\Delta}\sum_{\mathbf{i}\mathbf{j}\tilde{\mathbf{i}}\tilde{\mathbf{j}}} \Delta_{\mathbf{i}\tilde{\mathbf{i}}}^{(\rm t)*} \Delta_{\mathbf{j}\tilde{\mathbf{j}}}^{(\rm b)} t_{\textbf{i}\textbf{j}} t_{\tilde{\textbf{i}}\tilde{\textbf{j}}}+c.c.
\end{eqnarray}
Note that on the above Wick decomposition, we have only taken the parts relevant to the coupling between the pairing order parameters on the two mono-layers. Now if the pairing chiralities for $\Delta^{(\text{b})}_{\mathbf{j}\tilde{\mathbf{j}}}$ and $\Delta^{(\text{t})}_{\mathbf{i}\tilde{\mathbf{i}}}$ are opposite, we can, without lossing generality, let $\Delta^{(\text{b})}_{\mathbf{j}\tilde{\mathbf{j}}}\sim \Delta_{d+id}$ and $\Delta^{(\text{t})}_{\mathbf{i}\tilde{\mathbf{i}}}\sim \Delta_{d-id}$. By symmetry, we have
\begin{eqnarray}\label{symmetry}
\Delta^{(\text{b})}_{\hat P_{\frac{\pi}{3}}\mathbf{j}\hat P_{\frac{\pi}{3}}\tilde{\mathbf{j}}}=e^{i\frac{2\pi}{3}}\Delta^{(\text{b})}_{\mathbf{j}\tilde{\mathbf{j}}},\Delta^{(\text{t})}_{\hat P_{\frac{\pi}{3}}\mathbf{i}\hat P_{\frac{\pi}{3}}\tilde{\mathbf{i}}}=e^{-i\frac{2\pi}{3}}\Delta^{(\text{t})}_{\mathbf{i}\tilde{\mathbf{i}}},
t_{\hat P_{\frac{\pi}{3}}\mathbf{i}\hat P_{\frac{\pi}{3}}\mathbf{j}}=t_{\mathbf{i}\mathbf{j}},t_{\hat P_{\frac{\pi}{3}}\tilde{\mathbf{i}}\hat P_{\frac{\pi}{3}}\tilde{\mathbf{j}}}=t_{\tilde{\mathbf{i}}\tilde{\mathbf{j}}}.
\end{eqnarray}
Then from Eq. (\ref{Josephson1}), we have
\begin{eqnarray}\label{noncoupling}
F_{inter-layer} \approx -\frac{1}{2\Delta}\sum_{n=1}^6 e^{i\frac{4\pi}{3}n}{\sum_{\mathbf{i}}}'\sum_{\mathbf{j}\tilde{\mathbf{i}}\tilde{\mathbf{j}}} \Delta_{\mathbf{i}\tilde{\mathbf{i}}}^{(\rm t)*} \Delta_{\mathbf{j}\tilde{\mathbf{j}}}^{(\rm b)} t_{\textbf{i}\textbf{j}} t_{\tilde{\textbf{i}}\tilde{\textbf{j}}}+c.c. =0 +c.c.=0,
\end{eqnarray}
where ${\sum_{\mathbf{i}}}'$ represents the sum of one-sixth of all the sites $\mathbf{i}$. Eq. (\ref{noncoupling}) suggests that if the chiralities of the pairing order parameters from the two mono-layers are opposite, the system could not gain energy from the inter-layer Josephson coupling. Therefore, we only consider the case wherein the pairings in the two mono-layers are both $d+id$ or $d-id$.

When the pairing order parameters on both mono-layers of the QC-TBG are $d+id$, we can let them satisfy the following relation
\begin{eqnarray}
	\Delta_{d+id}^{(\rm b)}=\hat{P}_{\frac{\pi}{6}}\Delta_{d+id}^{(\rm t)},~~ \hat{P}_{\frac{\pi}{3}}\Delta_{d+id}^{(\rm \mu)}=e^{-i\frac{2\pi}{3}}\Delta_{d+id}^{(\rm \mu)},	
\end{eqnarray}
where $\mu=(\rm t,\rm b)$ is the layer index, $\hat{P}_\phi$ denotes the rotation by the angle $\phi$ and $\Delta^{(\text{t/b})}_{d+id}$ represent for the normalized gap form factors (either in the real- or $\mathbf{k}$ space) on the top/bottom layers. Setting the ``complex amplitudes'' of the gap functions on the top/bottom laters as $\psi_{\rm t/\rm b}$, the free energy function $F$ reads,
\begin{eqnarray}\label{fEdplusid}
F_{d+id}(\psi_{\rm t},\psi_{\rm b})=F_{0} (|\psi_{\rm t}|^2)+F_{0}(|\psi_{\rm b}|^2)-A(e^{i\theta}\psi_{\rm t}\psi_{\rm b}^*+c.c)+O(\psi^4).
\end{eqnarray}
Here by the phrase ``complex amplitudes'' we mean that the pairing gap functions on the t/b layers are $\psi_{\rm t/\rm b}\Delta_{d+id}^{(\rm t/\rm b)}$, where $\psi_{\rm t/\rm b}$ are the global amplitudes which are generally complex numbers. In Eq. (\ref{fEdplusid}), the $F_0-$ and the A- terms denote the contributions from each monolayer and their Josephson coupling respectively. The $A > 0$ and $\theta \in (-\pi,\pi]$ are real numbers. From the time-reversal symmetry, we know that when the pairing order parameters on both mono-layers of the QC-TBG are $d-id$, the free energy should be
\begin{eqnarray}\label{fEdminusid}
F_{d-id}(\psi_{\rm t},\psi_{\rm b})=F_{0}(|\psi_{\rm t}|^2)+F_{0}(|\psi_{\rm b}|^2)-A(e^{-i\theta}\psi_{\rm t}\psi_{\rm b}^*+c.c)+O(\psi^4)
\end{eqnarray}
In the following, we verify that Eq. (\ref{fEdplusid}) and Eq. (\ref{fEdminusid}) satisfy all symmetries of the system, including the time-reversal symmetry, the U(1)-gauge symmetry, and the point-group symmetry.

The time-reversal operation dictates
\begin{eqnarray}
	\psi_{(\rm t/\rm b)} \rightarrow \tilde\psi_{(\rm t/\rm b)}=	\psi_{(\rm t/\rm b)}^*,\Delta_{d+id}^{(\rm t/\rm b)} \rightarrow \Delta_{d+id}^{(\rm t/\rm b)*}=\Delta_{d-id}^{(\rm t/\rm b)}.
\end{eqnarray}
Therefore, the time-reversal symmetry of the system requires
\begin{eqnarray}\label{trs}
	F_{d+id}(\psi_{\rm t},\psi_{\rm b})=F_{d-id}(\psi_{\rm t}^*,\psi_{\rm b}^*)	
\end{eqnarray}
It can be seen that the free energy function given by Eq. (\ref{fEdplusid}) and Eq. (\ref{fEdminusid}) satisfies the time-reversal symmetry.

Similarly, the U(1)-gauge transformation dictates
\begin{eqnarray}
	\psi_{(\rm t/\rm b)} \rightarrow \tilde\psi_{(\rm t/\rm b)}=	e^{i\eta}\psi_{(\rm t/\rm b)},
\end{eqnarray}
where $\eta$ is an arbitrary phase angle. Therefore the global U(1)-gauge symmetry of the system requires,
\begin{eqnarray}\label{gt}
	F_{d\pm id}(\psi_{\rm t},\psi_{\rm b})=F_{d\pm id}(e^{i\eta}\psi_{\rm t},e^{i\eta}\psi_{\rm b})	
\end{eqnarray}
Because $\psi$ and $\psi^*$ always come in pair in Eq. (\ref{fEdplusid}) and Eq. (\ref{fEdminusid}), Eq. (\ref{gt}) is satisfied.

Finally, let's consider the point-group operations. We only need to consider the two generators of this group: one is the rotation by the angle $ \phi=\frac{\pi}{6}$, followed by a succeeding layer exchange, and the other can be chosen as the specular reflection operation $\sigma$ that changes the layer index. The former generator dictates
\begin{eqnarray}
\psi_{\rm t} \rightarrow \tilde\psi_{\rm t}=	e^{-i\frac{2\pi}{3}}\psi_{\rm b},\psi_{\rm b} \rightarrow \tilde\psi_{\rm b}=	\psi_{\rm t}.
\end{eqnarray}
The symmetry under this combined rotation and layer-exchange operation requires
\begin{eqnarray}\label{rs}
	F_{d+id}(\psi_{\rm t},\psi_{\rm b})=F_{d+id}(e^{-i\frac{2\pi}{3}}\psi_{\rm b},\psi_{\rm t}).
\end{eqnarray}
Eq.~\eqref{rs} is used in the main text to obtain $\theta=\frac{\pi}{3}$ or $\frac{-2\pi}{3}$, under which this equation is satisfied. The latter generator dictates
\begin{eqnarray}
	\psi_{\rm t} \rightarrow \tilde\psi_{\rm t}=	\psi_{\rm b}, \psi_{\rm b} \rightarrow \tilde\psi_{\rm b}=	\psi_{\rm t}, \Delta_{d+id}^{(\rm t/\rm b)} \rightarrow \Delta_{d-id}^{(\rm t/\rm b)}.
\end{eqnarray}
The symmetry under this specular reflection operation requires
\begin{eqnarray}\label{sr2}
	F_{d+id}(\psi_{\rm t},\psi_{\rm b})=F_{d-id}(\psi_{\rm b},\psi_{\rm t})	
\end{eqnarray}
Clearly, the free energy formula (\ref{fEdplusid}) and (\ref{fEdminusid}) satisfy Eq. (\ref{sr2}).

From the above analysis, we know that the free energy function provided by Eq. (\ref{fEdplusid}) and Eq. (\ref{fEdminusid}) satisfies all the symmetries of the system. Through minimization of the free energy function, we can obtain the $d\pm id$ or $g\pm ig$ pairing symmetries of the QC-TBG, as has already been provided in the main text.

\begin{figure}[htbp]
	\centering
	\includegraphics[width=0.95\textwidth]{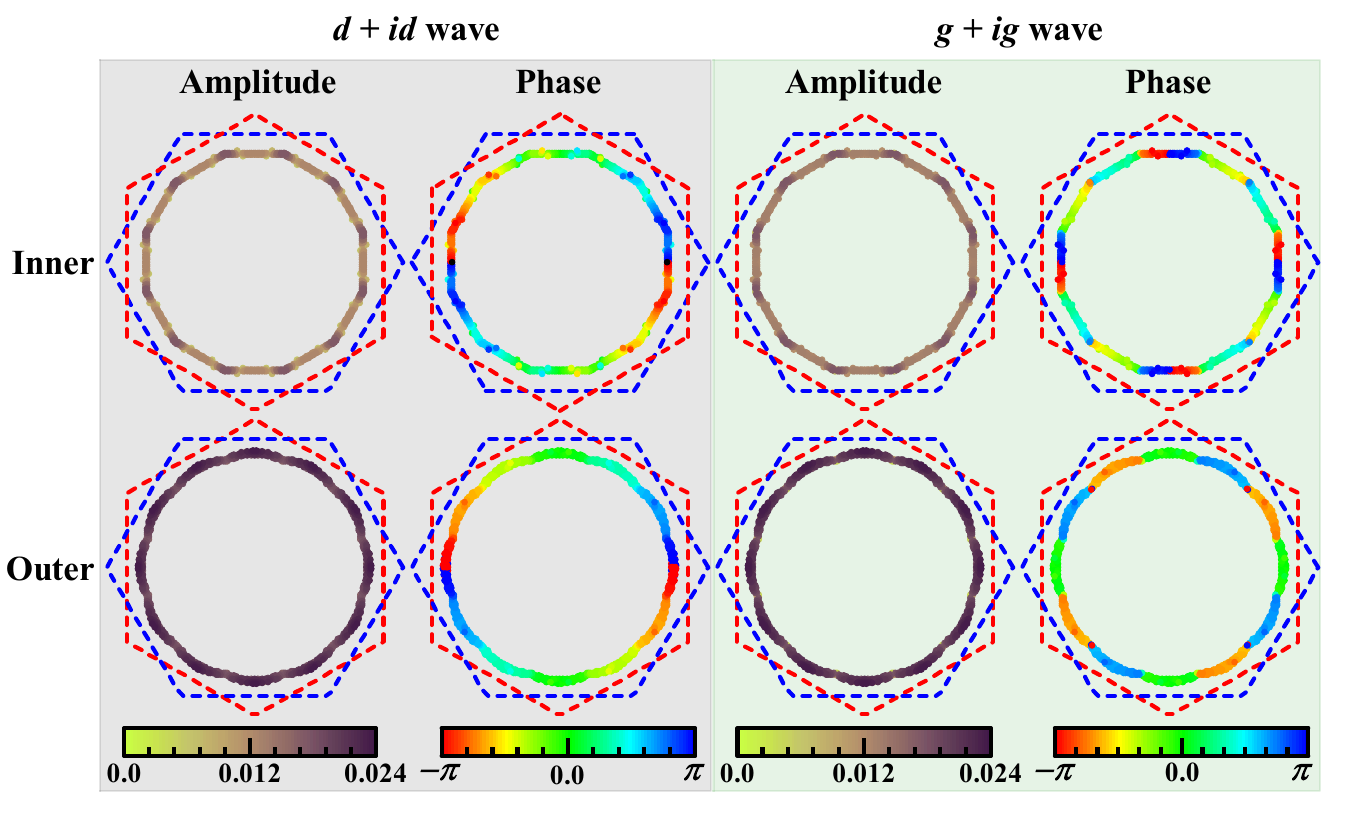}
	\caption{Distributions of the amplitude and phase of paring gap functions $d+id$- and $g+ig$-waves on the inner and outer Fermi surfaces when $\delta=0.32$. The figure shows the distributions of the gap functions within a narrow energy shell near the Fermi level.}
	\label{topo}
\end{figure}
\section{Topological properties}
This section provides detailed informations about the topological properties of the obtained $d+id$ and $g+ig$ TSCs, including their winding numbers, Chern numbers, the spontaneous super current and the Majorana edge states.

The distributions of the gap amplitudes of the obtained $d+id$- and $g+ig$-wave pairing gap functions for the typical doping level $\delta=0.32$ are shown on the two Fermi pockets separately in Fig.~\ref{topo}. It suggests that the two pairing states are fully gapped. This is due to that the gap nodal directions don't coincide with each other for the two basis functions of the degenerate $d$- or $g$-wave pairings, as shown in Fig.~\ref{dg}. This property provides the foundation for the two pairing states to be topologically nontrivial. We further plot the distributions of the phase angles of the complex $d+id$- and $g+ig$-wave pairing gap functions on the two Fermi pockets separately in Fig.~\ref{topo}. Obviously, the complex gap phase angle for the $d+id$-wave pairing changes $4\pi$ for each run around each pocket, leading to the winding number 2. The topological Chern number is equal to the summation of the winding numbers of each pocket~\cite{6, 7}. Considering the presence of two pockets, the Chern number should be $2\times2=4$. In contrast, the gap phase angle for the $g+ig$ changes $8\pi$ for each run around each pocket, leading to the winding number 4, and hence the Chern number $2\times4=8$.

The topological properties of the obtained TSCs are robust against slight deviation of the twist angle from $30\degree$ to, say 29.9$\degree$. Under such deviation, the point group decays to $D_6$, and our solution of Eq. (\ref{linear_eq}) at, say the doping level $\delta=0.32$, yields a leading pairing symmetry belonging to the 2D IRRP of $D_6$ with $L=2$. The $1:\pm i$ mixing of the two obtained degenerate basis functions leads to a distribution of the gap phase angles on the FS very similar as that of the $g+ig$-wave pairing shown in Fig.~\ref{topo}, yielding the same topological Chern numbers. The obtained state can be thought of as an approximate $g+ig$-wave TSC. The cases for other dopings are similar. Therefore, the pairing phase diagram for the case with the twist angle 29.9\degree is topologically the same as that for the case with the twist angle 30\degree.

The details for calculating the Majorana edge states are as follow. We transform Eq.~\eqref{HMF} into the real space and get the BdG Hamiltonian,
\begin{eqnarray}
H_{\rm MF} &=&
\sum_{\mathbf{ij}\sigma}(-t_{\mathbf{ij}}-\mu_{c}\delta_{\mathbf{ij}})
c^{\dagger}_{\mathbf{i}\sigma}c_{\mathbf{j}\sigma}+
\sum_{\mathbf{ij}}\left(\tilde{\Delta}_{\mathbf{ij}}c^{\dagger}_{\mathbf{i}\uparrow}c^{\dagger}_{\mathbf{j}\downarrow}+h.c.\right)\nonumber    \\
~&=&
\left(\begin{array}{cc}
\mathbf{c}_{\uparrow}^{\dagger} & \mathbf{c}_{\downarrow}\end{array}\right)\left(\begin{array}{cc}
-t-\mu_{c}I & \tilde{\Delta}\\
\text{\ensuremath{\tilde{\Delta}^{\dagger}}} & t+\mu_{c}I
\end{array}\right)\left(\begin{array}{c}
\mathbf{c}_{\uparrow}\\
\mathbf{c}_{\downarrow}^{\dagger}
\end{array}\right) \nonumber    \\
~&=&
\left(\begin{array}{cc}
\mathbf{c}_{\uparrow}^{\dagger} & \mathbf{c}_{\downarrow}\end{array}\right)
 H_{\rm BdG}
\left(\begin{array}{c}
\mathbf{c}_{\uparrow}\\
\mathbf{c}_{\downarrow}^{\dagger}
\end{array}\right) \nonumber    \\
~&=&
X^{\dagger}H_{\rm BdG}X = \sum_{m=1}^{2N_s}E_{m}\mathbf{\gamma}_{m}^{\dagger}\mathbf{\gamma}_{m}.
\end{eqnarray}
where $\mu_{c} $ is the chemical potential, $I$ is the identity matrix, $\mathbf{c}_\sigma=\left(c_{1\sigma}, c_{2\sigma}, \cdots\right),X^T=\left(\begin{array}{cc}
\mathbf{c}_{\uparrow} & \mathbf{c}_{\downarrow}^{\dagger}\end{array}\right),     X=\omega\mathbf{\gamma},  \\   \omega^{\dagger}H_{\rm BdG}\omega=diag(E_1,E_2,...,E_{2N_s})$,  and
\begin{equation}
\tilde{\Delta}_{\mathbf{ij}} = -\frac{1}{N}\sum_{\mathbf{k}\mu\alpha}\tilde\xi_{\mathbf{i},\mathbf{k}\mu\alpha}\tilde\xi^{*}_{\mathbf{j},\mathbf{k}\mu\alpha}\Delta_{\mathbf{k}\mu\alpha}.
\end{equation}
Under the open boundary condition, the eigenstates of $H_{\rm MF}$ with eigenvalue $E_{m}=0$ are the Majorana edge states, whose creation operators are
\begin{equation}
 \mathbf{\gamma}_{m}^{\dagger}=\sum_{i=1}^{N_s}\left(\omega_{i,m}\mathbf{c}_{i\uparrow}^{\dagger}+\omega_{i+N_s,m}\mathbf{c}_{i\downarrow}\right).
\end{equation}
where $N_s$ is the total site number, $\omega_{i,m}(\omega_{i+N_s,m})$ represents the particle(hole) part of the  Majorana edge states. In Fig. 3(d) of the main text, we show the probability distribution of the Majorana edge states in real space, i.e. $|\omega_{i,m}|^2+|\omega_{i+N,m}|^2$.

In the following, we provide the details for the calculation of the spontaneous super current. The $\alpha$-component ($\alpha=x,y$) of the vectorial current operator at site $\mathbf{i}$ is defined as
\begin{equation}
J_{\mathbf{i}\alpha}[\mathbf{A}]  =  -\frac{\delta H_{\rm MF}[\mathbf{A}]}{\delta A_{\mathbf{i}\alpha}} =  -\frac{\delta H_{\mathrm{TB}}[\mathbf{A}]}{\delta A_{\mathbf{i}\alpha}},\label{eq:current-operator-definition}
\end{equation}
where $\mathbf{A}$ is the vector potential with component $A_{\mathbf{i}\alpha}$. It appears in $\hat{H}_{\rm MF}[\mathbf{A}]$ through a modification of $\hat{H}_{\mathrm{TB}}$ into
\begin{eqnarray}\label{H_TB_A}
H_{\mathrm{TB}}[\mathbf{A}] & = & \ensuremath{-\sum_{\mathbf{ij}\sigma}t_{\mathbf{ij}}\exp({i\int_{\mathbf{i}}^{\mathbf{j}}}\mathbf{A}\cdot d\mathbf{l})c_{\mathbf{i}\sigma}^{\dagger}c_{\mathbf{j}\sigma}}.
\end{eqnarray}
For the study of the spontaneous super current, the field $\mathbf{A}$ goes to 0. In such a case, one gets \cite{x1},
\begin{eqnarray}\label{current_operator}
\lim_{\mathbf{A}\rightarrow0}J_{\mathbf{i}\alpha}[\mathbf{A}] & = & \frac{i}{2}\sum_{\mathbf{j}\sigma}t_{\mathbf{ij}}R_{\mathbf{ij},\alpha}c_{\mathbf{i}\sigma}^{\dagger} c_{\mathbf{j}\sigma}+h.c.,
\end{eqnarray}
where $R_{\mathbf{ij},\alpha}$ is the $\alpha$-component of $\mathbf{R}_{\mathbf{ij}}$. After some derivations, one can get
\begin{eqnarray}
J_{\mathbf{i}\alpha} & = & \frac{i}{2N}\sum_{\mathbf{j}\sigma}t_{\mathbf{ij}}R_{\mathbf{ij},\alpha}
\sum_{\mathbf{k}\mu\alpha\atop\mathbf{q}\nu\beta\sigma}\tilde\xi_{\mathbf{i},\mathbf{k}\mu\alpha}^{*}\tilde\xi_{\mathbf{j},\mathbf{q}\nu\beta}\tilde c_{\mathbf{k}\mu\alpha\sigma}^{\dagger}\tilde c_{\mathbf{q}\nu\beta\sigma}+h.c.\nonumber    \\
~&=&
\frac{i}{2N}\sum_{\mathbf{j}}t_{\mathbf{ij}}R_{\mathbf{ij},\alpha}\left(
\sum_{\mathbf{k}\mu\alpha\atop\mathbf{q}\nu\beta}
\tilde\xi_{\mathbf{i},\mathbf{k}\mu\alpha}^{*}\tilde\xi_{\mathbf{j},\mathbf{q}\nu\beta}\tilde c_{\mathbf{k}\mu\alpha\uparrow}^{\dagger}\tilde c_{\mathbf{q}\nu\beta\uparrow}+
\sum_{\mathbf{-k}\mu\alpha\atop\mathbf{-q}\nu\beta}
\tilde\xi_{\mathbf{i},\mathbf{k}\mu\alpha}\tilde\xi_{\mathbf{j},\mathbf{q}\nu\beta}^{*}\tilde c_{\mathbf{-k}\mu\alpha\downarrow}^{\dagger}\tilde c_{\mathbf{-q}\nu\beta\downarrow}\right)+h.c.
\end{eqnarray}
We prove below that the expectation values of the super currents on each site are always zero under the periodic boundary condition for each layer, that is
\begin{eqnarray}
\langle J_{\mathbf{i}\alpha}\rangle & = & \frac{i}{2N}\sum_{\mathbf{j}}t_{\mathbf{ij}}R_{\mathbf{ij},\alpha}
\left(\sum_{\mathbf{k}\alpha\mu}
\tilde\xi_{\mathbf{i},\mathbf{k}\mu\alpha}^{*}
\tilde\xi_{\mathbf{j},\mathbf{k}\mu\alpha}
\left\langle\tilde{ c}_{\mathbf{k}\mu\alpha\uparrow}^{\dagger}
\tilde c_{\mathbf{k}\mu\alpha\uparrow}\right\rangle
+\sum_{\mathbf{-k}\alpha\mu}
\tilde\xi_{\mathbf{i},\mathbf{k}\mu\alpha}
\tilde\xi_{\mathbf{j},\mathbf{k}\mu\alpha}^{*}
\left\langle\tilde c_{\mathbf{-k}\mu\alpha\downarrow}^{\dagger}\tilde c_{\mathbf{-k}\mu\alpha\downarrow}\right\rangle\right)+c.c.\nonumber    \\
~&=&
\frac{i}{2N}\sum_{\mathbf{j}}t_{\mathbf{ij}}R_{\mathbf{ij},\alpha}
\left(\sum_{\mathbf{k}\alpha\mu}
\tilde\xi_{\mathbf{i},\mathbf{k}\mu\alpha}^{*}
\tilde\xi_{\mathbf{j},\mathbf{k}\mu\alpha}v_{\mathbf{k}\mu\alpha}^{2}+
\sum_{\mathbf{-k}\alpha\mu}\tilde\xi_{\mathbf{i},\mathbf{k}\mu\alpha}
\tilde\xi_{\mathbf{j},\mathbf{k}\mu\alpha}^{*}v_{\mathbf{k}\mu\alpha}^{2}\right)+c.c.\nonumber    \\
~&=&
\frac{i}{N}\sum_{\mathbf{j}}t_{\mathbf{ij}}R_{\mathbf{ij},\alpha}\sum_{\mathbf{k}\alpha\mu}\mathbf{Re}
\left(\tilde\xi_{\mathbf{i},\mathbf{k}\mu\alpha}^{*}+\tilde\xi_{\mathbf{j},\mathbf{k}\mu\alpha}\right)v_{\mathbf{k}\mu\alpha}^{2}+c.c.\nonumber    \\
~&=&\frac{i}{N}\sum_{\mathbf{j}}t_{\mathbf{ij}}R_{\mathbf{ij},\alpha}\sum_{\mathbf{k}\alpha\mu}\mathbf{Re}
\left(\tilde\xi_{\mathbf{i},\mathbf{k}\mu\alpha}^{*}+\tilde\xi_{\mathbf{j},\mathbf{k}\mu\alpha}\right)v_{\mathbf{k}\mu\alpha}^{2}-\frac{i}{N}\sum_{\mathbf{j}}t_{\mathbf{ij}}R_{\mathbf{ij},\alpha}\sum_{\mathbf{k}\alpha\mu}\mathbf{Re}
\left(\tilde\xi_{\mathbf{i},\mathbf{k}\mu\alpha}^{*}+\tilde\xi_{\mathbf{j},\mathbf{k}\mu\alpha}\right)v_{\mathbf{k}\mu\alpha}^{2}\nonumber    \\ &=& 0,
\end{eqnarray}
with
\begin{equation}
v_{\mathbf{k}\mu\alpha}^2 = \frac{1}{2}\left(1-\frac{\tilde{\varepsilon}^{\mu\alpha}_\mathbf{k}-\mu_c}{\sqrt{\left|\tilde{\varepsilon}^{\mu\alpha}_\mathbf{k}-\mu_c\right|^2+\left|\Delta_{\mu\alpha}(\mathbf{k})\right|^2}}\right).
\end{equation}
Note that the intra-band pairing has been taken in the above derivation, which is usually physical. Therefore, the spontaneous super current cannot emerge under the periodic boundary condition for each layer, if only the intra-band pairing is considered. However, under the open boundary condition, the spontaneous super current can emerge at the edge.

Our numerical results reveal the existence of the Majorana edge states and spontaneous edge currents in the $d+id$ and $g+ig$ TSC states, with the latter case shown in the main text. These properties further justify the topological properties of the obtained TSCs.

\end{widetext}


\begin{thebibliography}{10}
\bibitem{footnote2} As the intra-layer hopping and interaction strengths are much stronger than the inter-layer ones, one can reasonably first let each two degenerate $d$-wave pairings on one mono-layer mix as $d\pm id$, and then consider their inter-layer coupling. Further more, we only consider the cases wherein the pairings from the two layers are both $d+id$ or $d-id$, because otherwise the system cannot gain energy from the inter-layer Josephson coupling. See the SM\cite{SM}.
\bibitem{footnote3} The pairing gap functions on the t/b layers are written as $\psi_{\text{t/b}}\Delta^{(\text{t/b})}_{d+id}$, where $\psi_{\text{t/b}}$ are complex-valued ``amplitudes''.
\end{thebibliography}
\end{document}